\documentclass[preprint,12pt, 3p]{elsarticle}
\usepackage[utf8]{inputenc}
\usepackage[T1]{fontenc} 

\usepackage{graphicx}
\usepackage[
    colorlinks=true,
    linkcolor=blue,
    filecolor=magenta,      
    urlcolor=cyan,
]{hyperref} 

\usepackage{float} 
\usepackage{xargs} 

\usepackage{dingbat}
\usepackage{color,soul}

\usepackage{longtable}
\usepackage{lipsum}  
\usepackage{adjustbox}

\usepackage{lineno}
\usepackage{amssymb}
\usepackage[figuresright]{rotating}
\usepackage{lipsum}
\usepackage{mathtools}
\usepackage{cuted}

\usepackage[pdftex,dvipsnames]{xcolor}  

\colorlet{soultransparent}{red!0}
\sethlcolor{soultransparent}

\usepackage[colorinlistoftodos,prependcaption,textsize=small]{todonotes}
\newcommandx{\unsure}[2][1=]{\todo[linecolor=red,backgroundcolor=red!25,bordercolor=red,#1]{#2}}
\newcommandx{\change}[2][1=]{\todo[linecolor=blue,backgroundcolor=blue!25,bordercolor=blue,#1]{#2}}
\newcommandx{\info}[2][1=]{\todo[linecolor=OliveGreen,backgroundcolor=OliveGreen!25,bordercolor=OliveGreen,#1]{#2}}
\newcommandx{\improvement}[2][1=]{\todo[linecolor=Plum,backgroundcolor=Plum!25,bordercolor=Plum,#1]{#2}}
\newcommandx{\thiswillnotshow}[2][1=]{\todo[disable,#1]{#2}}

\usepackage{natbib}

\usepackage[automake]{glossaries}

\usepackage{glossary-mcols}
\setglossarystyle{mcolindex}
\makeglossaries

\usepackage{subcaption}

\newacronym{ml}{ML}{Machine Learning}
\newacronym{dl}{DL}{Deep Learning}
\newacronym{ai}{AI}{Artificial Intelligence}
\newacronym[plural=CNNs]{cnn}{CNN}{Convolutional Neural Network}
\newacronym[plural=DBNs]{dbn}{DBN}{Deep Belief Network}
\newacronym{lstm}{LSTM}{Long-Short Term Memory}
\newacronym{crps}{CRPS}{Continuous Ranked Probability Score}
\newacronym[plural=RNNs]{rnn}{RNN}{Recurrent Neural Network}
\newacronym[plural=ANNs]{ann}{ANN}{Artificial Neural Network}
\newacronym[plural=DNNs]{dnn}{DNN}{Deep Neural Network}
\newacronym{mlp}{MLP}{Multilayer Perceptron}
\newacronym[plural=DMLPs]{dmlp}{DMLP}{Deep Multilayer Perceptron}
\newacronym[plural=AEs]{ae}{AE}{Autoencoder}
\newacronym{hci}{HCI}{Human-Computer Interaction}
\newacronym[plural=GPs]{gp}{GP}{Genetic Programming}
\newacronym[plural=GAs]{ga}{GA}{Genetic Algorithm}
\newacronym[plural=ECs]{ec}{EC}{Evolutionary Computation}
\newacronym[plural=MOEAs]{moea}{MOEA}{Multiobjective Evolutionary Algorithm}
\newacronym{relu}{ReLU}{Rectified Linear Unit}
\newacronym{sgd}{SGD}{Stochastic Gradient Descent}
\newacronym{adagrad}{AdaGrad}{Adaptive Gradient Algorithm}
\newacronym{rmsprop}{RMSProp}{Root Mean Square Propagation}
\newacronym{adam}{ADAM}{Adaptive Moment Estimation}
\newacronym{cd}{CD}{Contrastive Divergence}
\newacronym{kldivergence}{KL-Divergence}{Kullback Leibler Divergence}
\newacronym{ais}{AIS}{Annealed Importance Sampling}
\newacronym{rl}{RL}{Reinforcement Learning}
\newacronym{dp}{DP}{Dynamic Programming}
\newacronym{mc}{MC}{Monte Carlo}
\newacronym{td}{TD}{Temporal Difference}
\newacronym{dwnn}{DWNN}{Deep and Wide Neural Network}
\newacronym{srnn}{SRNN}{Stacked Recurrent Neural Network}
\newacronym{arma}{ARMA}{Autoregressive Moving Average}
\newacronym{elm}{ELM}{Extreme Learning Machine}
\newacronym{gbt}{GBT}{Gradient Boosted Trees}
\newacronym{ti}{TI}{Technical Indicator}
\newacronym{gan-fd}{GAN-FD}{GAN for minimizing Forecast error loss and Direction prediction loss}
\newacronym{rcnn}{RCNN}{Recurrent CNN}
\newacronym{ar}{AR}{Autoregressive}
\newacronym{rf}{RF}{Random Forest}
\newacronym{sp500}{S\&P500}{Standard’s \& Poor’s 500 Index}
\newacronym{nifty}{NIFTY}{National Stock Exchange of India}
\newacronym{sse}{SSE}{Shanghai Stock Exchange}
\newacronym{hsi}{HSI}{Hong Kong Hang Seng Index}
\newacronym{taiex}{TAIEX}{Taiwan Capitalization Weighted Stock Index}
\newacronym{dow30}{DOW30}{Dow Jones Industrial Average 30}
\newacronym{kospi}{KOSPI}{The Korea Composite Stock Price Index}
\newacronym{vxn}{VXN}{NASDAQ100 Volatility Index}
\newacronym{bovespa}{Bovespa}{Brazilian Stock Exchange}
\newacronym{omx}{OMX}{Stockholm Stock Exchange}
\newacronym{egarch}{EGARCH}{Exponential GARCH}
\newacronym{bi-lstm}{Bi-LSTM}{Bidirectional LSTM}
\newacronym{har}{HAR}{Heterogeneous Autoregressive Process}
\newacronym{gasvr}{GASVR}{\acrshort{ga} with a \acrshort{svr}}
\newacronym{flann}{FLANN}{Functional Link Neural Network}
\newacronym{ea}{EA}{Evolutionary Algorithm}
\newacronym{lda}{LDA}{Latent Dirichlet Allocation}
\newacronym{cdbn}{CDBN}{Continuous-valued Deep Belief Networks}
\newacronym{crbm}{CRBM}{Continuous Restricted Boltzman machine}
\newacronym{pnn}{PNN}{Probabilistic Neural Network}
\newacronym{xgboost}{XGBoost}{eXtreme Gradient Boosting}
\newacronym{hmm}{HMM}{Hidden Markov Model}
\newacronym{tgru}{TGRU}{Two-stream GRU}
\newacronym{har-gasvr}{HAR-GASVR}{\acrshort{har} with a \acrshort{gasvr}}
\newacronym{rmdn}{RMDN}{Recurrent Mixture Density Network}
\newacronym{rmdn-garch}{RMDN-GARCH}{\acrshort{rmdn} with a \acrshort{garch}}
\newacronym{rw}{RW}{Random Walk}
\newacronym{mrs}{MRS}{Markov Regime Switching}
\newacronym{williamr}{William\%R}{Williams Percent Range}
\newacronym{pposc}{PPOSC}{Percentage Price Oscillator} 
\newacronym{gml}{GML}{Generalized Linear Model}
\newacronym{dfnn}{DFNN}{Deep Feedforward Neural Network}
\newacronym{ffnn}{FFNN}{Feedforward Neural Network}
\newacronym{nn}{NN}{Neural Network}
\newacronym{aar}{AAR}{Annual Rate of Return}
\newacronym{ac}{AC}{Autocorrelation}
\newacronym{amex}{AMEX}{American Stock Exchange}
\newacronym{areturn}{AR}{Active Return}
\newacronym{arch}{ARCH}{Autoregressive Conditional Heteroscedasticity}
\newacronym{arima}{ARIMA}{Autoregressive Integrated Moving Average}
\newacronym{atr}{ATR}{Average True Range}
\newacronym{auc}{AUC}{Area Under the Curve}
\newacronym{auroc}{AUROC}{Area Under the Receiver Operating Characteristics}
\newacronym{ba}{BA}{Balanced Accuracy}
\newacronym{belm}{BELM}{Basic Extreme Learning Machine}
\newacronym{betc}{BETC}{Break Even Transaction Cost}
\newacronym{bist}{BIST}{Istanbul Stock Exchange Index}
\newacronym{bi-gru}{Bi-GRU}{Bidirectional Gated Recurrent Unit}
\newacronym{boll}{BOLL}{Bollinger Band}
\newacronym{bp}{BP}{Backpropagation}
\newacronym{bptt}{BPTT}{Backpropagation Through Time}
\newacronym{bse}{BSE}{Bombay Stock Exchange}
\newacronym{cagr}{CAGR}{Compound Annual Growth Rate}
\newacronym{car}{CAR}{Cumulative Abnormal Return}
\newacronym{cart}{CART}{Classification and Regression Trees}
\newacronym{cc}{CC}{Correlation Coefficient}
\newacronym{cci}{CCI}{Commodity Channel Index}
\newacronym{cdax}{CDAX}{German Stock Market Index Calculated by Deutsche Börse}
\newacronym{cdbn-fg}{CDBN-FG}{Fuzzy Granulation with Continuous-valued Deep Belief Networks}
\newacronym{cds}{CDS}{Credit Default Swaps}
\newacronym{cew}{CEW}{Emerging Markets Currency Index}
\newacronym{cgan}{CGAN}{Conditional \acrshort{gan}}
\newacronym{cme}{CME}{Chicago Mercantile Exchange}
\newacronym{coefficient}{coefficient}{}
\newacronym{crsp}{CRSP}{Center for Research in Security Prices}
\newacronym{cse}{CSE}{Colombo Stock Exchange}
\newacronym{csi}{CSI}{China Securities Index}
\newacronym{cwn}{cWN}{Conditional Wavenet}
\newacronym{da}{DA}{Direction Accuracy}
\newacronym{dax}{DAX}{The Deutscher Aktienindex}
\newacronym{dcnn}{DCNN}{Deep Convolutional Neural Network}
\newacronym{ddpg}{DDPG}{Deep Deterministic Policy Gradient}
\newacronym{de}{DE}{Differential Evolution} 
\newacronym{deep-fasp}{Deep-FASP}{The Financial Aspect and Sentiment Prediction task with Deep neural networks}
\newacronym{deepcnl}{DeepCNL}{Deep Co-investment Network Learning}
\newacronym{dffn}{DFFN}{Deep Feed Forward Network}
\newacronym{dgm}{DGM}{Deep Neural Generative Model}
\newacronym{djia}{DJIA}{Dow Jones Industrial Average}
\newacronym{dlr}{DLR}{Deep Learning Representation}
\newacronym{dmi}{DMI}{Directional Movement Index}
\newacronym{dof}{DOF}{Degrees of Freedom}
\newacronym{dpa}{DPA}{Direction Prediction Accuracy}
\newacronym{dql}{DQL}{Deep Q-Learning}
\newacronym{drl}{DRL}{Deep Reinforcement Learning}
\newacronym{drse}{DRSE}{Deep Random Subspace Ensembles}
\newacronym{dtw}{DTW}{Dynamic Time Warping}
\newacronym{ema}{EMA}{Exponential Moving Average}
\newacronym{emd2fnn}{EMD2FNN}{Empirical Mode Decomposition and Factorization Machine based Neural Network}
\newacronym{etf}{ETF}{Exchange-Traded Fund}
\newacronym{far}{FAR}{False Acceptance Rate}
\newacronym{fddr}{FDDR}{Fuzzy Deep Direct Reinforcement Learning}
\newacronym{fe-qar}{FE-QAR}{Fixed Effects Quantile VAR}
\newacronym{fiqa}{FiQA}{Financial Opinion Mining and Question Answering Challange}
\newacronym{fn}{FN}{False Negative}
\newacronym{fnn}{FNN}{Fully Connected Neural Network}
\newacronym{fcnn}{FCNN}{Fully Connected Neural Network}
\newacronym{fp}{FP}{False Positive}
\newacronym{fpe}{FPE}{Akaike’s Minimum Final Prediction Error}
\newacronym{fpga}{FPGA}{Field Programmable Gate Array}
\newacronym{frr}{FRR}{False Rejection Rate}
\newacronym{ftse}{FTSE}{London Financial Times Stock Exchange Index}
\newacronym{g-mean}{G-mean}{Geometric Mean}
\newacronym{gaf}{GAF}{Gramian Angular Field}
\newacronym{gan}{GAN}{Generative Adversarial Network}
\newacronym{garch}{GARCH}{Generalised Auto-Regressive Conditional Heteroscedasticity}
\newacronym{gbdt}{GBDT}{Gradient-Boosted-DecisionTrees}
\newacronym{glm}{GLM}{Generalized Linear Model}
\newacronym{gpu}{GPU}{Graphic Processing Unit}
\newacronym{gru}{GRU}{Gated-Recurrent Unit}
\newacronym{gspc}{GSPC}{S\&P500 Commodity Price Index}
\newacronym{han}{HAN}{Hybrid Attention Network}
\newacronym{hft}{HFT}{High Frequency Trading}
\newacronym{hit}{HIT}{Hit Rate}
\newacronym{hmrpso}{HMRPSO}{Modified Version of PSO}
\newacronym{hs}{HS}{China Shanghai Shenzhen Stock Index}
\newacronym{ibb}{IBB}{iShares Nasdaq Biotechnology ETF}
\newacronym{ic}{IC}{Information Coeffiencient}
\newacronym{ir}{IR}{Information Ratio}
\newacronym{ise100}{ISE100}{Istanbul Stock Exchange Index}
\newacronym{ixic}{IXIC}{NASDAQ Composite Index}
\newacronym{kelm}{KELM}{Kernel Extreme Learning Machine}
\newacronym{ks}{KS}{Kolmogorov–Smirnov}
\newacronym{lar}{LAR}{Linear Auto-regression Predictor}
\newacronym{lfm}{LFM}{Lookahead Factor Models}
\newacronym{lob}{LOB}{Limit Order Book Data}
\newacronym{lrnfis}{LRNFIS}{Locally Recurrent Neuro-fuzzy Information System}
\newacronym{ma}{MA}{Moving Average}
\newacronym{macd}{MACD}{Moving Average Convergence and Divergence}
\newacronym{mad}{MAD}{Mean Absolute Deviation}
\newacronym{madr}{MADR}{Moving Average Deviation Rate}
\newacronym{mae}{MAE}{Mean Absolute Error}
\newacronym{mam}{MAM}{Moving Average Mapping}
\newacronym{map}{MAP}{Maximum Absolute Percentage Error}
\newacronym{mape}{MAPE}{Mean Absolute Percentage Error}
\newacronym{mar}{MAR}{Mean Abnormal Return}
\newacronym{mase}{MASE}{Mean Standard Deviation}
\newacronym{mcc}{MCC}{Matthew Correlation Coefficient}
\newacronym{mda}{MDA}{Multilinear Discriminant Analysis}
\newacronym{mdd}{MDD}{Maximum Drawdown}
\newacronym{mdp}{MDP}{Markov Decision Process}
\newacronym{mfi}{MFI}{Money Flow Index}
\newacronym{mi}{MI}{Mutual Information}
\newacronym{modrl}{MODRL}{Multi-objective Deep Reinforcement Learning}
\newacronym{moe}{MoE}{Mixture of Experts}
\newacronym{mse}{MSE}{Mean Squared Error}
\newacronym{msfe}{MSFE}{Mean Squared Forecast Error}
\newacronym{mspe}{MSPE}{Mean Squared Prediction Error}
\newacronym{mtm}{MTM}{Momentum}
\newacronym{narmax}{NARMAX}{Nonlinear Autoregressive Moving Average model with exogenous inputs}
\newacronym{nasdaq}{NASDAQ}{National Association of Securities Dealers Automated Quotations}
\newacronym{nes}{NES}{Natural Evolution Strategies}
\newacronym{nikkei}{NIKKEI}{Tokyo Nikkei Index}
\newacronym{nlp}{NLP}{Natural Language Processing}
\newacronym{nmae}{NMAE}{Normalized Mean Absolute Error}
\newacronym{nmse}{NMSE}{Normalized Mean Square Error}
\newacronym{nymex}{NYMEX}{New York Mercantile Exchange}
\newacronym{nyse}{NYSE}{New York Stock Exchange}
\newacronym{obv}{OBV}{On Balance Volume}
\newacronym{ochl}{OCHL}{Open,Close,High, Low}
\newacronym{ochlv}{OCHLV}{Open,Close,High, Low, Volume}
\newacronym{pca}{PCA}{Principal Component Analysis}
\newacronym{pcc}{PCC}{Pearson’s Correlation Coefficient}
\newacronym{pcd}{PCD}{Percentage of Correct Direction}
\newacronym{plr}{PLR}{Piecewise Linear Representation}
\newacronym{pocid}{POCID}{Percentage of Change in Direction}
\newacronym{ppo}{PPO}{Proximal Policy Optimization}
\newacronym{profit}{PROFIT}{Average Annual Profit of the Model}
\newacronym{psn}{PSN}{Psi-Sigma Network}
\newacronym{pso}{PSO}{Particle Swarm Optimization}
\newacronym{r-sq}{R$^2$}{Squared correlation, Non-linear regression multiple correlation}
\newacronym{r1}{r1}{Correlation coefficient between actual value and prediction value}
\newacronym{r2}{r2}{Correlation coefficient between actual return and prediction return}
\newacronym{ra}{RA}{Rolling Average}
\newacronym{raf}{RAF}{Random Forests}
\newacronym{rbf}{RBF}{Radial Basis Function Neural Network}
\newacronym{rbm}{RBM}{Restricted Boltzmann Machine}
\newacronym{rceflann}{RCEFLANN}{Recurrent Computationally Efficient Functional Link Neural Network}
\newacronym{rci}{RCI}{Rank Correlation Index}
\newacronym{return}{RETURN}{Average Annual Returns of the Model}
\newacronym{rmse}{RMSE}{Root Mean Square Error}
\newacronym{rmsre}{RMSRE}{Root Mean Square Relative Error}
\newacronym{roa}{ROA}{Return on Assets}
\newacronym{roc}{ROC}{Price of Change}
\newacronym{rse}{RSE}{Relative Squared Error}
\newacronym{rsi}{RSI}{Relative Strength Index}
\newacronym{sae}{SAE}{Stacked Autoencoder}
\newacronym{sar}{SAR}{Parabolic Stop and Reverse}
\newacronym{sci}{SCI}{SSE Composite Index}
\newacronym{sd}{SD}{Standard Deviation (also referred as the Greek letter r)}
\newacronym{sdae}{SDAE}{Stacked Denoising Autoencoders}
\newacronym{sfm}{SFM}{State Frequency Memory}
\newacronym{si}{SI}{Stochastic Index}
\newacronym{slp}{SLP}{Single Layer Perceptron}
\newacronym{smape}{SMAPE}{Symmetric Mean Absolute Percentage Error}
\newacronym{som}{SOM}{Self-Organising Map}
\newacronym{sr}{SR}{Sharpe-ratio}
\newacronym{svd}{SVD}{Singular Value Decomposition}
\newacronym{svm}{SVM}{Support Vector Machine}
\newacronym{svr}{SVR}{Support Vector Regressor}
\newacronym{szse}{SZSE}{Shenzhen Stock Exchange Composite Index}
\newacronym{talib}{TALIB}{Technical Analysis Library Package}
\newacronym{tar}{TAR}{Threshold Autoregressive}
\newacronym{vec}{VEC}{Vector Error Correction model}
\newacronym{rhe}{RHE}{Recurrent Hybrid Elman}
\newacronym{tdnn}{TDNN}{Timedelay Neural Network}
\newacronym{theil-u}{THEIL-U}{Theil's inequality coefficient}
\newacronym{tn}{TN}{True Negative}
\newacronym{tp}{TP}{True Positive}
\newacronym{tr}{TR}{Total Return}
\newacronym{tse}{TSE}{Tokyo Stock Exchange}
\newacronym{tunindex}{TUNINDEX}{Tunisian Stock Market Index}
\newacronym{twse}{TWSE}{Taiwan Stock Exchange}
\newacronym{uwn}{uWN}{Unconditional WaveNet}
\newacronym{var}{VAR}{Vector Auto Regression}
\newacronym{vix}{VIX}{S\&P500 Volatility Index}
\newacronym{vr}{VR}{Variance Reduction}
\newacronym{vwl}{VWL}{WL Kernel-based Method}
\newacronym{vxd}{VXD}{Dow Jones Industrial Average Volatility Index}
\newacronym{wba}{WBA}{Weighted Balanced Accuracy}
\newacronym{weka}{WEKA}{Waikato Environment for Knowledge Analysis}
\newacronym{whr}{WHR}{Weighted Hit Rate}
\newacronym{wmtr}{WMTR}{Weighted Multichannel Time-series Regression}
\newacronym{wpr}{WPR}{William \% R}
\newacronym{wsurt}{WSURT}{Wilcoxon Sum-rank Test}
\newacronym{wt}{WT}{Wavelet Transforms}
\newacronym{true}{TRUE}{True Range of Price Movements}
\newacronym{nse}{NSE}{National Stock Exchange of India}
\newacronym{norm-rmse}{norm-RMSE}{Normalized \acrshort{rmse}}
\newacronym{taq}{TAQ}{Trade and Quote}
\newacronym{hr}{HR}{Hit Rate}
\newacronym{std}{STD}{Standard Deviation}
\newacronym{ise}{ISE}{Istanbul Stock Exchange Index}
\newacronym{gdax}{GDAX}{Global Digital Asset Exchange}
\newacronym{wti}{WTI}{West Texas Intermediate}
\newacronym{mm}{MM}{Markov Model}
\newacronym{hmae}{HMAE}{Heteroscedasticity Adjusted MAE}
\newacronym{hmse}{HMSE}{Heteroscedasticity Adjusted MSE}
\newacronym{spy}{SPY}{SPDR S\&P 500 ETF}
\newacronym{ssec}{SSEC}{Shanghai Stock Exchange Composite}
\newacronym{kse}{KSE}{Korea Stock Exchange}
\newacronym{ibovespa}{IBOVESPA}{Indice Bolsa de Valores de Sao Paulo}
\newacronym{dji}{DJI}{Dow Jones Index}
\newacronym{tfidf}{TF-IDF}{Term Frequency-Inverse Document Frequency}
\newacronym{lr}{LR}{Logistic Regression}
\newacronym{tema}{TEMA}{Triple Exponential Moving Average}
\newacronym{b-h}{B\&H}{Buy and Hold}
\newacronym{wcn}{WCN}{Wavenet Convolution Network} 
\newacronym{fhs}{FHS}{Firefly Harmony Search} 

\newacronym{manualsearch}{MS}{Manual Search}
\newacronym{gridsearch}{GS}{Grid Search}
\newacronym{randomsearch}{RS}{RandomSearch}
\newacronym{smbgo}{SMBGO}{Sequential Model-Based Global Optimization}
\newacronym{gpa}{GPA}{The Gaussian Process Approach} 
\newacronym{tspea}{TSPEA}{Tree-structured Parzen Estimator Approach}

\newacronym{cae}{CAE}{Convolutional Autoencoder}

\newacronym{bhc}{BHC}{Bank Holding Companies}
\newacronym{dpower}{DP}{Discriminant Power}
\newacronym{crix}{CRIX}{The Cryptocurrency Index}
\newacronym{mvn}{MVN}{Multivariate Normal Distribution}
\newacronym{mv-t}{MV-t}{Multivariate t Distribution}

\newacronym{dcnl}{DCNL}{Deep Co-investment Network Learning}
\newacronym[longplural={Deep Gaussian Processes}]{dgp}{DGP}{Deep Gaussian Process}

\begin{document}

\begin{frontmatter}


\title{Deep Learning for Financial Applications : A Survey}

\author[addr]{Ahmet Murat Ozbayoglu}
\author[addr]{Mehmet Ugur Gudelek}
\author[addr]{Omer Berat Sezer}

\address[addr]{Department of  Computer Engineering, TOBB University of Economics and Technology, Ankara, Turkey}

\begin{abstract}
Computational intelligence in finance has been a very popular topic for both \hl{academia} and financial industry in the last few decades. Numerous studies have been published resulting in various models. Meanwhile, within the \gls{ml} field, \gls{dl} started getting a lot of attention recently, mostly due to its outperformance over the classical models. Lots of different implementations of \gls{dl} exist today, and the broad interest is continuing. Finance is one particular area where \gls{dl} models started getting traction, however, the playfield is wide open, a lot of research opportunities still exist. In this paper, we tried to provide a state-of-the-art snapshot of the developed \gls{dl} models for financial applications, as of today. We not only categorized the works according to their intended subfield in finance but also analyzed them based on their \gls{dl} models. In addition, we also aimed at identifying possible future implementations and highlighted the pathway for the ongoing research within the field. 
\end{abstract}


\end{frontmatter}

\section{Introduction}
\label{sec:introduction}

Stock market forecasting, algorithmic trading, credit risk assessment, portfolio allocation, asset pricing and derivatives market are among the areas where \gls{ml} researchers focused on developing models that can provide real-time working solutions for the financial industry. Hence, a lot of publications and implementations exist in the literature. 

However, within the \gls{ml} field, \gls{dl} is an emerging area with a rising interest every year. As a result, an increasing number of \gls{dl} models for finance started appearing in conferences and journals. Our focus in this paper is to present different implementations of the developed financial \gls{dl} models in such a way that the researchers and practitioners that are interested in the topic can decide which path they should take. 

In this paper, we tried to provide answers to the following research questions:
\begin{itemize}
\item {What financial application areas are of interest to \gls{dl} community?}
\item {How mature is the existing research in each of these application areas?}
\item {What are the areas that have promising potentials from an academic/industrial research perspective?}
\item {Which \gls{dl} models are preferred (and more successful) in different applications?}
\item {How do \gls{dl} models pare against traditional soft computing / \gls{ml} techniques?}
\item {What is the future direction for \gls{dl} research in Finance?}
\end{itemize}

Our focus was solely on \gls{dl} implementations for financial applications. A substantial portion of the computational intelligence for finance research is devoted to financial time series forecasting. However, we preferred to concentrate on those studies in a separate survey paper \cite{Sezer_2019a} in order to be able to pinpoint other, less covered application areas. Meanwhile, we decided to include algorithmic trading studies with \gls{dl} based trading strategies which may or may not have an embedded time series forecasting component. 

For our search methodology, we surveyed and carefully reviewed the studies that came to our attention from the following sources: \hl{ScienceDirect, ACM Digital Library, Google Scholar, arXiv.org, ResearchGate, Google keyword search for DL and finance}

The range of our survey spanned not only journals and conferences, but also Masters and PhD theses, book chapters, arXiv papers and noteworthy technical papers that came up in Google searches. Furthermore, we \hl{only chose the} articles that were written in English. It is worth to mention that we encountered a few studies that were written in a different language, but had English abstracts. However, for overall consistency, we decided not to include those studies in our survey. 

Most of the papers in this survey used the term ``deep learning" in their model description and they were published in the last 5 years. However, we also included some older papers that implemented deep learning models even though they were not called ``deep learning" models at their time of publication. Some examples for such models include \gls{rnn}, Jordan-Elman networks. 

To best of our knowledge, this will be the first comprehensive ``deep learning for financial applications" survey paper. As will be introduced in the next section, a lot of \gls{ml} surveys exist for different areas of finance, however, no study has concentrated on \gls{dl} implementations. We genuinely believe our study will highlight the major advancements in the field and provide a roadway for the intended researchers that would like to develop \gls{dl} models for different financial application areas.

The rest of the paper is structured as follows. After this brief introduction, in Section~\ref{sec:ml_in_finance}, the existing surveys that are focused on \gls{ml} and soft computing studies for financial applications are presented. In Section~\ref{sec:dl}, we will provide the basic working \gls{dl} models that are used in finance, i.e. \gls{cnn}, \gls{lstm}, etc. Section~\ref{sec:financial_application} will focus on the implementation areas of the \gls{dl} models in finance. Some of these include algorithmic trading, credit risk assessment, portfolio allocation, asset pricing, fraud detection and derivatives market. After briefly stating the problem definition in each subsection, \gls{dl} implementations of each associated problem will be given. 

In Section~\ref{sec:snapshot}, these studies will be compared and some overall statistical results will be presented including histograms about the yearly distribution of different subfields, models, publication types, etc. These statistics will not only demonstrate the current state for the field but also will show which areas are mature, which areas still have opportunities and which areas are getting accelerated attention.
Section~\ref{sec:discussion} will have discussions about what has been done in the field so far and where the industry is going. The chapter will also include the achievements and expectations of both academia and the industry. Also, open areas and recommended research topics will be mentioned. Finally, in Section~\ref{sec:conclusions}, we will summarize the findings and \hl{conclude.}

\section{Machine Learning in Finance }
\label{sec:ml_in_finance}

Finance has always been one of the most studied application areas for \gls{ml}, starting as early as 40 years ago. So far, thousands of research papers were published in various fields within finance, and the overall interest does not seem to diminish anytime soon. Even though this survey paper is solely focused on \gls{dl} implementations, we wanted to provide the audience with some insights about previous \gls{ml} studies by citing the related surveys within the last 20 years.

There are a number of \gls{ml} surveys and books with a general perspective such that they do not concentrate on any particular implementation area. The following survey papers fall into that category. Bahrammirzaee et al. \cite{Bahrammirzaee_2010} compared \glspl{ann}, Expert Systems and Hybrid models for various financial applications. Zhang et al. \cite{Zhang_2004} reviewed the data mining techniques including \gls{ga}, rule-based systems, \glspl{nn} preferred in different financial application areas. Similarly, Mochn et al. \cite{Mochn_2007} also provided insights about financial implementations based on soft computing techniques like fuzzy logic, probabilistic reasoning and \glspl{nn}. Even though Pulakkazhy et al. \cite{Pulakkazhy_2013} focused particularly on data mining models in banking applications, they still had a span of several subtopics within the field. Meanwhile, Mullainathan et al. \cite{Mullainathan_2017} studied the \gls{ml} implementations from a high level and econometric point of view. Likewise, Gai et al. \cite{Gai_2018} reviewed the Fintech studies and implementations not only from an \gls{ml} perspective but in general. The publications in \cite{Kovalerchuk_2000, Aliev_2004, Brabazon_2008, Dymowa_2011} constitute some of the books that cover the implementations of soft computing models in finance. 

Meanwhile, there are some survey papers that are also not application area-specific but rather focused on particular \gls{ml} techniques. One of those soft computing techniques is the family of \glspl{ea}, i.e. \gls{ga}, \gls{pso}, etc. commonly used in financial optimization implementations like Portfolio Selection. Chen et al. \cite{Chen_2002s} wrote a book covering \glspl{ga} and \gls{gp} in Computational Finance. Later, Castillo et al. \cite{Castillo_Tapia_2007}, Ponsich et al. \cite{Ponsich_2013}, Aguilar-Rivera et al. \cite{Aguilar_Rivera_2015} extensively surveyed \glspl{moea} on portfolio optimization and other various financial applications. 

Since \glspl{ann} were quite popular among researchers, a number of survey papers were just dedicated to them. Wong et al. \cite{Wong_1998} covered early implementations of \glspl{ann} in finance. Li et al. \cite{Li_2010} reviewed implementations of \glspl{ann} for stock price forecasting and some other financial applications. Lately, Elmsili et al. \cite{Elmsili_2018} contained \gls{ann} applications in economics and management research in their survey.

In addition, LeBaron \cite{LeBaron_2006} covered the studies focused on agent-based computational finance. Meanwhile, Chalup et al. \cite{Chalup} wrote a book chapter on kernel methods in financial applications which includes models like \gls{pca}, \gls{svm}.

And then, there are application-specific survey papers that single out particular financial areas which are quite useful and informative for researchers that already know what they are looking for. These papers will be covered in the appropriate subsections of Section \ref{sec:financial_application} during problem description. In the next section, brief working structures of the \gls{dl} models used in the financial applications will be given.

\section{Deep Learning}
\label{sec:dl}

Deep Learning is a particular type of \gls{ml} that consists of multiple \gls{ann} layers. It provides high-level abstraction for data modelling \cite{LeCun2015}. In the literature, different \gls{dl} models exist: \gls{dmlp}, \gls{cnn}, \gls{rnn}, \gls{lstm}, \glspl{rbm}, \glspl{dbn}, and \glspl{ae}.

\subsection{Deep Multi Layer Perceptron (DMLP)}

In the literature, \gls{dmlp} was the first proposed \gls{ann} model of its kind. \gls{dmlp} networks consist of input, output and hidden layers just like an ordinary \gls{mlp}; however, the number of layers in \gls{dmlp} is more than \gls{mlp}. Each neuron in every layer has input (x), weight (w) and bias (b) terms. An output of a neuron in the neural network is illustrated in Equation~\ref{eq:neuron1}. In addition, each neuron has a nonlinear activation function which produces the output of that neuron through accumulating weighted inputs from the neurons in the preceding layer. Sigmoid \cite{Cybenko_1989}, hyperbolic tangent   \cite{Kalman_1992}, \gls{relu} \cite{Nair_2010}, leaky \gls{relu} \cite{Maas_2013}, swish \cite{Ramachandran_2017}, and softmax\cite{Goodfellow-et-al-2016} are among the most preferred nonlinear activation functions in the literature.

\begin{equation}\label{eq:neuron1} 
y_i= \sigma (\sum\limits_{i} W_i x_i + b_i) 
\end{equation}

\begin{figure}[H]
\centering
\includegraphics[width=6in]{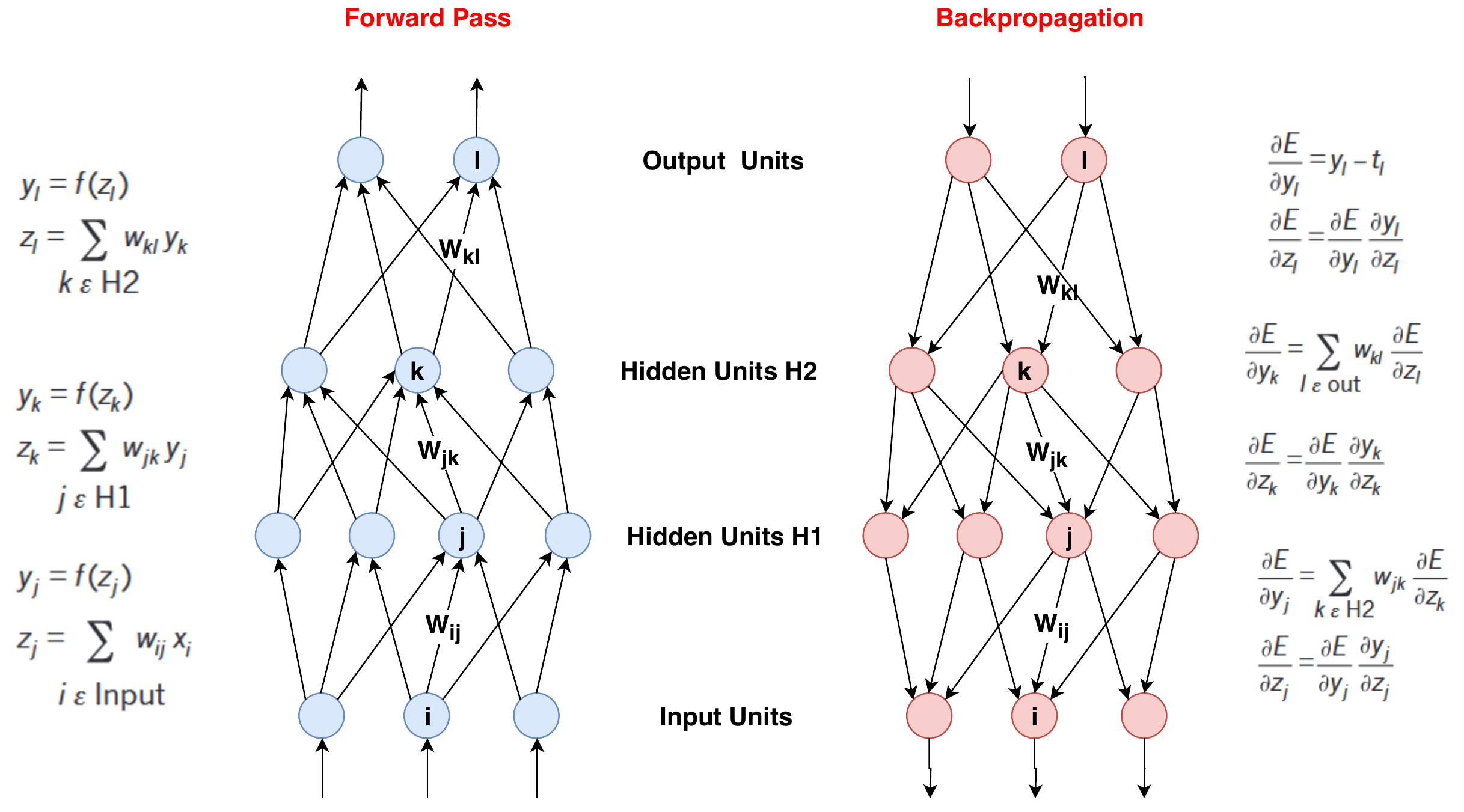}
\caption{Deep Multi Layer Neural Network Forward Pass and Backpropagation \cite{LeCun2015}}
\label{fig:forward-backward}
\end{figure}

With multi-layer deep \glspl{ann}, more efficient classification and regression performances are achieved when compared against shallow nets. \glspl{dmlp}' learning process is implemented through backpropagation. The amount of the output error in the output layer neurons is also reflected back to the neurons in the previous layers. In \gls{dmlp}, \gls{sgd} method is (mostly) used for the optimization of learning (to update the weights of the connections between the layers). In Figure~\ref{fig:forward-backward}, a \gls{dmlp} model, the layers, the neurons in layers, the weights between the neurons are shown.

\subsection{Convolutional Neural Networks (CNNs)}

\begin{figure}[H]
\centering
\includegraphics[width=6.4in]{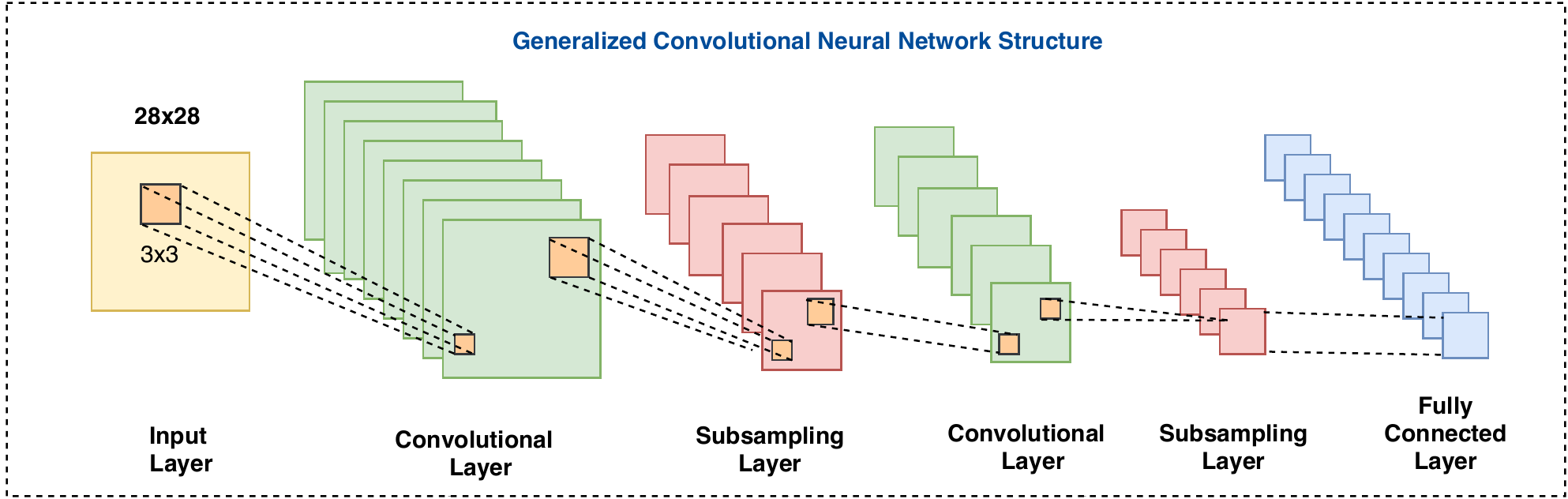}
\caption{Generalized Convolutional Neural Network Architecture}
\label{fig:cnn_architecture}
\end{figure}

\gls{cnn} is a type of \gls{dnn} that is mostly used for image classification, image recognition problems. In its methodology, the whole image is scanned with filters. In the literature, 1x1, 3x3 and 5x5 filter sizes are mostly used. In most of the \gls{cnn} architectures, there are different types of layers: convolutional, pooling (average or maximum), fully connected layers. \gls{cnn} consists of convolutional layers based on the convolutional operation. Figure~\ref{fig:cnn_architecture} shows the generalized CNN architecture that has different layers: convolutional, subsampling (pooling), fully connected layers.

\subsection{Recurrent Neural Network (RNN)}

In the literature, \gls{rnn} \hl{has been mostly} used on sequential data such as time-series data, audio and speech data, language. It consists of \gls{rnn} units that are structured consecutively. Unlike feed-forward networks, \glspl{rnn} use internal memory to process the incoming inputs. \glspl{rnn} are used in the analysis of the time series data in various fields (handwriting recognition, speech recognition, etc).

There are different types of \gls{rnn} structures: one to many, many to one, many to many. Generally, \gls{rnn} processes the input sequence series one by one at a time, during its operation. Units in the hidden layer hold information about the history of the input in the "state vector" \cite{LeCun2015}.  \glspl{rnn} can be trained using the \gls{bptt} method. Using \gls{bptt}, the differentiation of the loss at any time $t$ has reflected the weights of the network at the previous time. Training of \glspl{rnn} are more difficult than \glspl{ffnn} and the training period of \glspl{rnn} takes longer. 

\begin{figure}[H]
\centering
\includegraphics[width=3.8in]{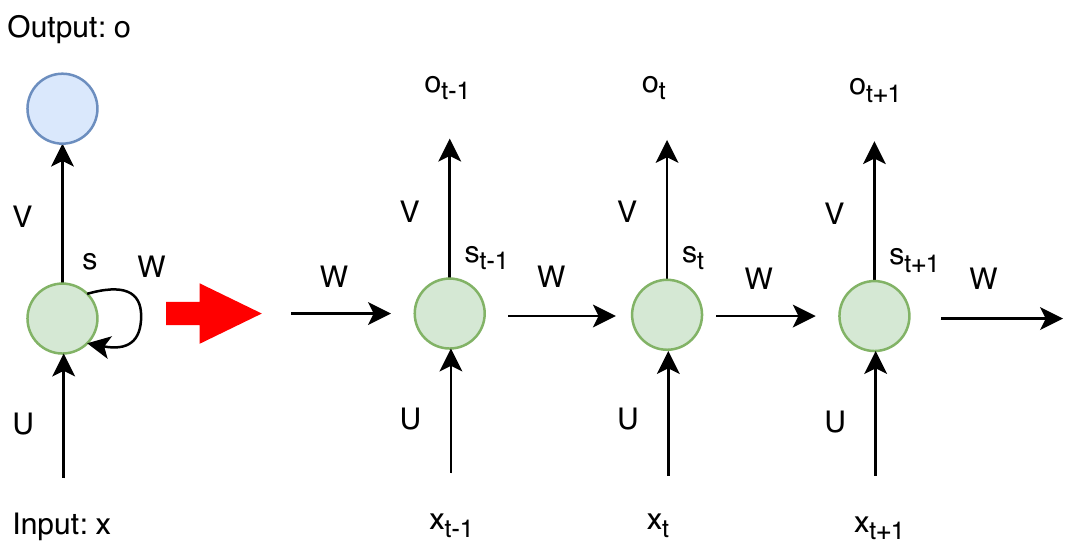}
\caption{RNN cell through time\cite{LeCun2015}}
\label{fig:rnn_fig2}
\end{figure}

In Figure~\ref{fig:rnn_fig2}, the information flow in the \gls{rnn}'s hidden layer is divided into discrete times. The status of the node S at different times of $t$ is shown as $s_t$, the input value $x$ at different times is $x_t$, and the output value $o$ at different times is shown as $o_t$. The parameter values ($U, W, V$) are always used in the same step. 

\subsection{Long Short Term Memory (LSTM)}

\begin{figure}[H]
\centering
\includegraphics[width=3.8in]{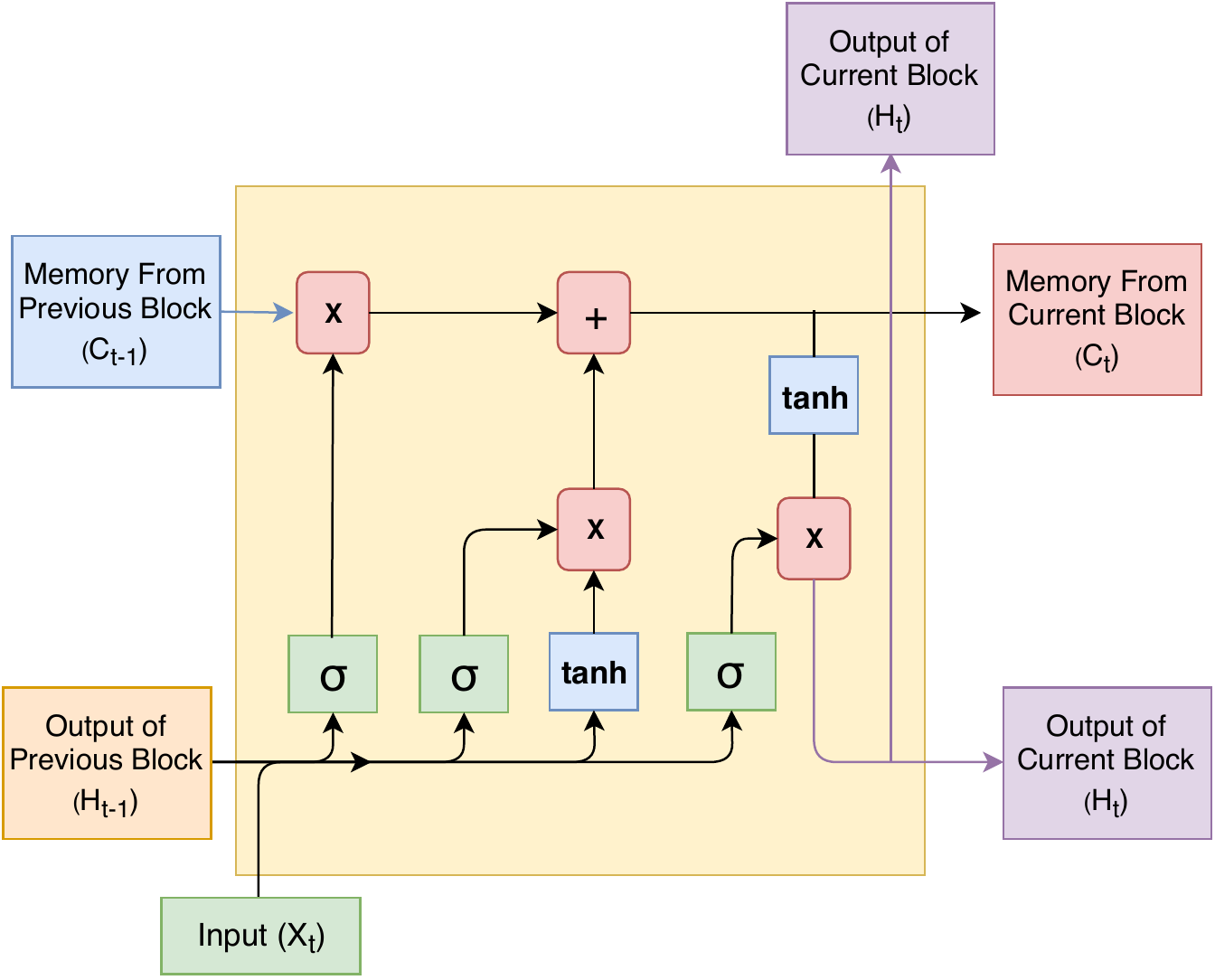}
\caption{Basic LSTM Unit \cite{hochreiter1997lstm}}
\label{fig:lstm}
\end{figure}

\gls{lstm} network \cite{hochreiter1997lstm} is a different type of \gls{dl} network specifically intended for sequential data analysis. The advantage of \gls{lstm} networks lies in the fact that both short term and long term values in the network can be remembered. Therefore,  \gls{lstm} networks are mostly used for sequential data analysis (automatic speech recognition, language translation, handwritten character recognition, time-series data forecasting, etc.) by \gls{dl} researchers. \gls{lstm} networks consist of \gls{lstm} units. \gls{lstm} unit is composed of cells having input, output and forget gates. These three gates regulate the information flow. With these features,  each cell remembers the desired values over arbitrary time intervals. \gls{lstm} cells combine to form layers of neural networks. Figure~\ref{fig:lstm} illustrates the basic LSTM unit ($\sigma_g$: sigmoid function, $tanh$: hyperbolic tangent function, $X$: multiplication, $+$: addition).

\subsection{Restricted Boltzmann Machines (RBMs)}

\begin{figure}[H]
\centering
\includegraphics[width= 3.7in]{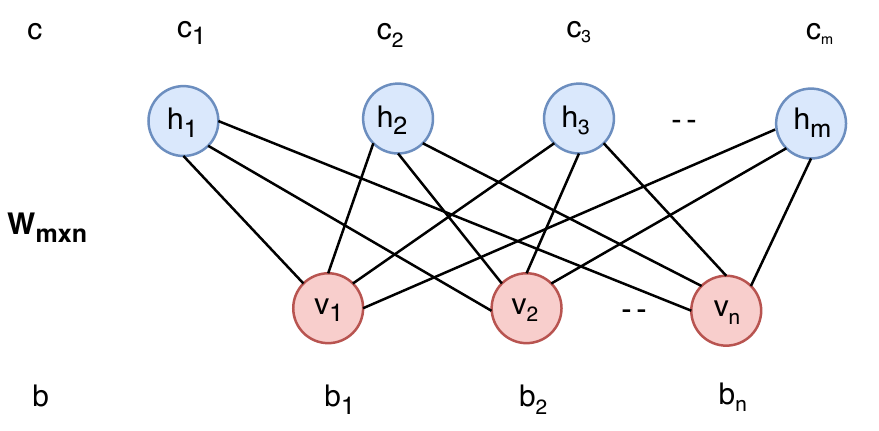}
\caption{RBM Visible and Hidden Layers \cite{Qiu2014}}
\label{fig:rbm_fig}
\end{figure}

\gls{rbm} is a different type of \gls{ann} model that can learn the probability distribution of the input set \cite{Qiu2014}. \glspl{rbm} are mostly used for dimensionality reduction, classification, and feature learning. \gls{rbm} is a bipartite, undirected graphical model that consists of two layers; visible and hidden layer.  The units in the layer are not connected to each other. Each cell is a computational point that processes the input. Each unit makes stochastic decisions about whether transmitting the input data or not. The inputs are multiplied by specific weights, certain threshold values (bias) are added to the input values, then the calculated values are passed through an activation function. In the reconstruction stage, the results in the outputs re-enter the network as the input, then they exit from the visible layer as the output. The values of the previous input and the values after the processes are compared. The purpose of the comparison is to reduce the difference. The learning is performed multiple times on the network \cite{Qiu2014}. \gls{rbm} is a two-layer, bipartite, and undirected graphical model that consists of two layers; visible and hidden layers (Figure~\ref{fig:rbm_fig}).  The layers are not connected among themselves. The disadvantage of \gls{rbm} is its tricky training. ``\glspl{rbm} are tricky because although there are good estimators of the log-likelihood gradient, there are no known cheap ways of estimating the log-likelihood itself" \cite{Bengio_2012}.

\subsection{Deep Belief Networks (DBNs)}

\begin{figure}[H]
\centering
\includegraphics[width=2.6in]{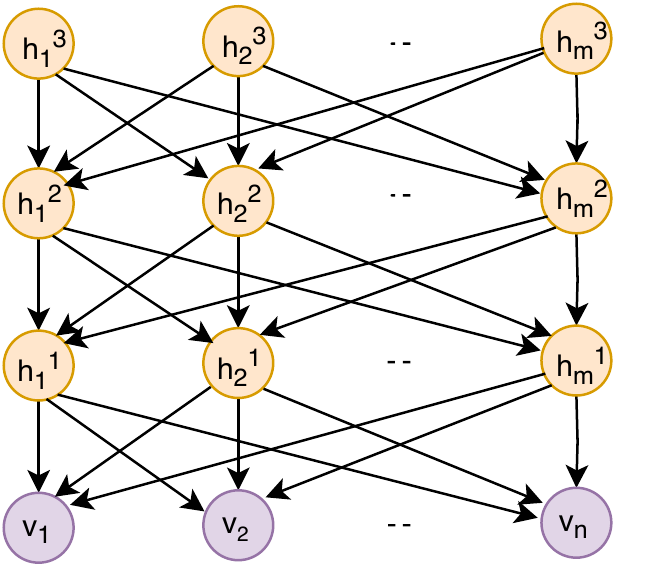}
\caption{Deep Belief Network \cite{Qiu2014} }
\label{fig:deep_belief_fig}
\end{figure}

\gls{dbn} is a type of \gls{ann} that consists of a stack of \gls{rbm} layers. \gls{dbn} is a probabilistic generative model that consists of latent variables. \glspl{dbn} are used for finding independent and discriminative features in the input set using an unsupervised approach. \gls{dbn} can learn to reconstruct the input set in a probabilistic way during the training process. Then the layers on the network begin to detect the discriminative features. After the learning step, supervised learning is carried out to perform for the classification \cite{Hinton2006}. Figure~\ref{fig:deep_belief_fig} illustrates the \gls{dbn} structure.

\subsection{Autoencoders (AEs)}

\gls{ae} networks are commonly used in \gls{dl} models, wherein they remap the inputs (features) such that the inputs are more representative for the classification. In other words, \gls{ae} networks perform an unsupervised feature learning process. A representation of a data set is learned by reducing the dimensionality with an \gls{ae}. In the literature, \glspl{ae} have been used for feature extraction and dimensionality reduction \cite{Goodfellow-et-al-2016,Vincent_2008}. The architecture of an \gls{ae} has similarities with that of a \gls{ffnn}. It consists of an input layer, output layer and one (or more) hidden layer that connects them together. The number of nodes in the input layer and the number of nodes in the output layer are equal to each other in \glspl{ae}, and they have a symmetrical structure. \glspl{ae} contain two components: encoder and decoder.

\begin{figure}[H]
\centering
\includegraphics[width=3.84in]{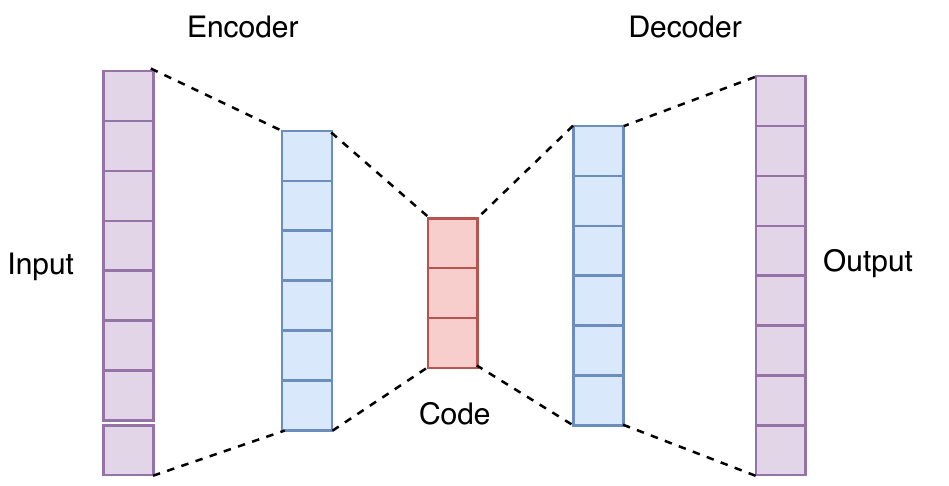}
\caption{Basic Autoencoder Structure }
\label{fig:ae}
\end{figure}

The advantages of the usage of  \gls{ae} are dimensionality reduction and feature learning. However, reducing dimensions and feature extraction in \gls{ae} cause some drawbacks. Focusing on minimizing the loss of the data relationship in the code of \gls{ae} causes the loss of some significant data relationship. This may be a drawback of \gls{ae}\cite{Meng_2017}. Figure~\ref{fig:ae} shows the basic \gls{ae} structure.

\subsection{Other Deep Structures}
\hl{The }\gls{dl} \hl{models are not limited to the ones mentioned in the previous subsections. Some of the other well-known structures that exist in the literature are } \gls{drl}, \glspl{gan}\hl{, Capsule Networks, }\glspl{dgp}\hl{. Meanwhile, to the best of our knowledge, we have not encountered any noteworthy academic or industrial publication on financial applications using these models so far, with the exception of } \gls{drl} \hl{ which started getting attention lately. However, that does not imply that these models do not fit well with the financial domain. On the contrary, they offer great potentials for researchers and practitioners participating in finance and deep learning community who are willing to go the extra mile to come up with novel solutions.}

Since research for model developments in \gls{dl} is ongoing, new structures keep on coming. However, the aforementioned models currently cover almost all of the published work. Next section will provide details about the implementation areas along with the preferred \gls{dl} models.

\section{Financial Applications}
\label{sec:financial_application}

There are a lot of financial applications of soft computing in the literature. \gls{dl} has been studied in most of them, although, some opportunities still exist in a number of fields. 

Throughout this section, we categorized the implementation areas and presented them in separate subsections. Besides, in each subsection we tabulated the representative features of the relevant studies in order to provide as much information as possible in the limited space.

Also, the readers should note that there were some overlaps between different implementation areas for some papers. There were two main reasons for that: In some papers, multiple problems were addressed separately, for e.g. text mining was studied for feature extraction, then algorithmic trading was implemented. For some other cases, the paper might fit directly into multiple implementation areas due to the survey structure, for e.g. cryptocurrency portfolio management. In such cases we included the papers in all of the relevant subsections creating some overlaps.

Some of the existing study areas can be grouped as follows:

\subsection{Algorithmic Trading}

Algorithmic trading (or Algo-trading) is defined as buy-sell decisions made solely by algorithmic models. These decisions can be based on some simple rules, mathematical models, optimized processes, or as in the case of machine/deep learning, highly complex function approximation techniques. With the introduction of electronic online trading platforms and frameworks, algorithmic trading took over the finance industry in the last two decades. As a result, Algo-trading models based on \gls{dl} also started getting attention.

Most of the Algo-trading applications are coupled with price prediction models for market timing purposes. As a result, a majority of the price or trend forecasting models that trigger buy-sell signals based on their prediction are also considered as Algo-trading systems. However, there are also some studies that propose stand-alone Algo-trading models focused on the dynamics of the transaction itself by optimizing trading parameters such as bid-ask spread, analysis of limit order book, position-sizing, etc. \gls{hft} researchers are particularly interested in this area. Hence, \gls{dl} models also started appearing in \gls{hft} studies.

Before diving into the \gls{dl} implementations, it would be beneficial to briefly mention about the existing \gls{ml} surveys on Algo-trading. Hu et al. \cite{Hu_2015} reviewed the implementations of various \glspl{ea} on Algorithmic Trading Models. Since financial time series forecasting is highly coupled with algorithmic trading, there are a number of \gls{ml} survey papers focused on Algo-trading models based on forecasting. The interested readers can refer to \cite{Sezer_2019a} for more information. 

As far as the \gls{dl} research is concerned, Table~\ref{table:algorithmic_trading_1}, Table~\ref{table:algorithmic_trading_2}, and Table~\ref{table:algorithmic_trading_3} present the past and current status of algo-trading studies based on \gls{dl} models. The papers are distributed to these tables as follows: Table~\ref{table:algorithmic_trading_1} has the particular algorithmic trading implementations that are embedded with time series forecasting models, whereas Table~\ref{table:algorithmic_trading_2} is focused on classification based (Buy-sell Signal, or Trend Detection) algo-trading models. Finally, Table~\ref{table:algorithmic_trading_3} presents stand-alone studies or other algorithmic trading models (pairs trading, arbitrage, etc) that do not fit into the above clustering criteria.

Most of the Algo-trading studies were concentrated on the prediction of stock or index prices. Meanwhile, \gls{lstm} was the most preferred \gls{dl} model in these implementations. In \cite{Karaoglu_2017}, market microstructures based trade indicators were used as the input into \gls{rnn} with Graves \gls{lstm} to perform the price prediction for algorithmic stock trading. Bao et al. \cite{Bao_2017} used technical indicators as the input into \gls{wt}, \gls{lstm} and \glspl{sae} for the forecasting of stock prices. In \cite{Liu_2017}, \gls{cnn} and \gls{lstm} \hl{model structures} were implemented together (\gls{cnn} was used for stock selection, \gls{lstm} was used for price prediction). 

\begingroup
\footnotesize
\fontsize{7}{9}\selectfont

\begin{longtable}{
                p{0.03\linewidth}
                p{0.18\linewidth}
                p{0.08\linewidth}
                p{0.16\linewidth}
                p{0.14\linewidth}
                p{0.15\linewidth}
                p{0.10\linewidth}
                }
            
\caption{Algo-trading Applications Embedded with Time Series Forecasting Models} \\

\hline
\textbf{Art.}                                                          & 
\textbf{Data Set}                                                       & 
\textbf{Period}                                                    & 
\textbf{Feature Set}                                                    & 
\textbf{Method}                                                 & 
\textbf{Performance Criteria}                                           & 
\textbf{Environment}                                                      \\ 
\hline
\endhead 

\\

\cite{Karaoglu_2017} & GarantiBank in \acrshort{bist}, Turkey & 2016 & \acrshort{ochlv},  Spread, Volatility, Turnover, etc. & \acrshort{plr}, Graves \acrshort{lstm} & \acrshort{mse}, \acrshort{rmse}, \acrshort{mae}, \acrshort{rse}, Correlation R-square & Spark \\ \hline
\cite{Bao_2017} & \acrshort{csi}300, Nifty50, \acrshort{hsi}, Nikkei 225, \acrshort{sp500}, \acrshort{djia} & 2010-2016 & \acrshort{ochlv}, Technical Indicators & \acrshort{wt}, Stacked autoencoders, \acrshort{lstm} & \acrshort{mape}, Correlation coefficient, \acrshort{theil-u} & - \\ \hline
\cite{Liu_2017} & Chinese Stocks & 2007-2017 & \acrshort{ochlv} & \acrshort{cnn} + \acrshort{lstm} & Annualized Return, Mxm Retracement & Python  \\ \hline
\cite{Zhang_2017} & 50 stocks from \acrshort{nyse} & 2007-2016 & Price data & \acrshort{sfm} & \acrshort{mse} & - \\ \hline
\cite{Tran_2017} & The LOB of 5 stocks of Finnish Stock Market & 2010 & FI-2010 dataset: bid/ask and volume & \acrshort{wmtr}, \acrshort{mda} & Accuracy, Precision, Recall, F1-Score & - \\ \hline
\cite{Deng_2017} & 300 stocks from \acrshort{szse}, Commodity & 2014-2015 & Price data & \acrshort{fddr}, \acrshort{dnn}+\acrshort{rl} & Profit, return, \acrshort{sr}, profit-loss curves & Keras \\ \hline
\cite{Fischer_2018} & \acrshort{sp500} Index  & 1989-2005 & Price data, Volume & \acrshort{lstm} & Return, \acrshort{std}, \acrshort{sr}, Accuracy & Python, TensorFlow, Keras, R, H2O \\ \hline
\cite{Mourelatos_2018} & Stock of National Bank of Greece (ETE). & 2009-2014 & \acrshort{ftse}100, \acrshort{djia}, GDAX, \acrshort{nikkei}225, EUR/USD, Gold & \acrshort{gasvr}, \acrshort{lstm} & Return, volatility, \acrshort{sr}, Accuracy & Tensorflow \\ \hline
\cite{Si_2017} & Chinese stock-IF-IH-IC contract & 2016-2017 & Decisions for price change & \acrshort{modrl}+\acrshort{lstm} & Profit and loss, \acrshort{sr} & - \\ \hline
\cite{Yong_2017} & Singapore Stock Market Index & 2010-2017 & \acrshort{ochl} of last 10 days of Index & \acrshort{dnn} & \acrshort{rmse}, \acrshort{mape}, Profit, \acrshort{sr} & - \\ \hline
\cite{Lu_2017} & GBP/USD  & 2017 & Price data & Reinforcement Learning + \acrshort{lstm} + \acrshort{nes} & \acrshort{sr}, downside deviation ratio, total profit & Python, Keras, Tensorflow \\ \hline
\cite{Dixon_2016} & Commodity, FX future, \acrshort{etf} & 1991-2014 & Price Data & \acrshort{dnn} & \acrshort{sr}, capability ratio, return & C++, Python \\ \hline
\cite{Korczak_2017} & USD/GBP, \acrshort{sp500}, \acrshort{ftse}100, oil, gold & 2016 & Price data & \acrshort{ae} + \acrshort{cnn} & \acrshort{sr}, \% volatility, avg return/trans, rate of return & H2O \\ \hline
\cite{Spilak_2018} & Bitcoin, Dash, Ripple, Monero, Litecoin, Dogecoin, Nxt, Namecoin & 2014-2017 & MA, BOLL, the CRIX returns, Euribor interest rates, \acrshort{ochlv} & \acrshort{lstm}, \acrshort{rnn}, \acrshort{mlp} & Accuracy, F1-measure & Python, Tensorflow \\ \hline
\cite{Jeong_2019} & \acrshort{sp500}, \acrshort{kospi}, \acrshort{hsi}, and EuroStoxx50 & 1987-2017 & 200-days stock price & Deep Q-Learning, \acrshort{dnn} & Total profit,  Correlation & - \\ \hline
\cite{Krauss_2017} & Stocks in the \acrshort{sp500} & 1990-2015 & Price data & \acrshort{dnn}, \acrshort{gbt}, \acrshort{rf} & Mean return, \acrshort{mdd}, Calmar ratio & H2O \\ \hline
\cite{GooglePatent} & Fundamental and Technical Data, Economic Data & - & Fundamental , technical and market information & \acrshort{cnn} & - & - \\ \hline

\label{table:algorithmic_trading_1}
\end{longtable}
\endgroup

Using a different model, Zhang et. al. \cite{Zhang_2017} proposed a novel \gls{sfm} recurrent network for stock price prediction with multiple frequency trading patterns and achieved better prediction and trading performances. In an \gls{hft} trading system, Tran et al. \cite{Tran_2017} developed a \gls{dl} model that implements price change forecasting through mid-price prediction using high-frequency limit order book data with tensor representation. In \cite{Deng_2017}, the authors used \gls{fddr} for stock price prediction and trading signal generation.

For index prediction, the following studies are noteworthy. In \cite{Fischer_2018}, the price prediction of S\&P500 index using \gls{lstm} was implemented. Mourelatos et al. \cite{Mourelatos_2018} compared the performance of \gls{lstm} and \gls{gasvr} for Greek Stock Exchange Index prediction.  Si et al. \cite{Si_2017} implemented Chinese intraday futures market trading model with \gls{drl} and \gls{lstm}. Yong et al. \cite{Yong_2017} used feed-forward \gls{dnn} method and \gls{ochl} of the time series index data to predict Singapore Stock Market index data. 

Forex or cryptocurrency trading was implemented in some studies. In \cite{Lu_2017}, agent inspired trading using deep (recurrent) reinforcement learning and \gls{lstm} was implemented and tested on the trading of GBP/USD.  In \cite{Dixon_2016}, feedforward deep \gls{mlp} was implemented for the prediction of commodities and FX trading prices. Korczak et al. \cite{Korczak_2017} implemented a forex trading (GBP/PLN) model using several different input parameters on a multi-agent-based trading environment. One of the agents was using \gls{cnn} as the prediction model and outperformed all other models.

On the cryptocurrency side, Spilak et al. \cite{Spilak_2018} used several cryptocurrencies (Bitcoin, Dash, Ripple, Monero, Litecoin, Dogecoin, Nxt, Namecoin) to construct a dynamic portfolio using \gls{lstm}, \gls{rnn}, \gls{mlp} methods. 

In a versatile study, Jeong et al. \cite{Jeong_2019} combined deep Q-learning and \gls{dnn} to implement price forecasting and they intended to solve three separate problems: Increasing profit in a market, prediction of the number of shares to trade, and preventing overfitting with insufficient financial data.  

In \cite{Sezer_2017}, technical analysis indicator's (\gls{rsi}) buy \& sell limits were optimized with \gls{ga} which was used for buy-sell signals. After optimization, \gls{dmlp} was also used for function approximation. In \cite{Navon_2017}, the authors combined deep \gls{fnn} with a selective trade strategy unit to predict the next price. In \cite{Troiano_2018}, the crossover and \gls{macd} signals were used to predict the trend of the Dow 30 stocks' prices. Sirignano et al. \cite{Sirignano_2018} proposed a novel method that used limit order book flow and history information for the determination of the stock movements using \gls{lstm} model.  Tsantekidis et al. \cite{Tsantekidis_2017} also used limit order book time series data and \gls{lstm} method for the trend prediction.

Several studies focused on utilizing \gls{cnn} based models due to their success in image classification problems. However, in order to do that, the financial input data needed to be transformed into images which required some creative preprocessing. Gudelek et al. \cite{Gudelek_2017} converted time series of price data to 2-dimensional images using technical analysis and classified them with deep \gls{cnn}. Similarly, Sezer et al. \cite{Sezer_2018} also proposed a novel technique that converts financial time series data that consisted of technical analysis indicator outputs to 2-dimensional images and classified these images using \gls{cnn} to determine the trading signals. In \cite{Hu_2018_a}, candlestick chart graphs were converted into 2-dimensional images. Then, unsupervised convolutional \gls{ae} was fed with the images to implement portfolio construction. Tsantekidis et al. \cite{Tsantekidis_2017_a} proposed a novel method that used the last 100 entries from the limit order book to create a 2-dimensional image for the stock price prediction using \gls{cnn} method. In \cite{Gunduz_2017}, an innovative method was proposed that uses \gls{cnn} with correlated features combined together to predict the trend of the stocks prices. Finally, Sezer et al. \cite{Sezer_2019} directly used bar chart images as inputs to \gls{cnn} and predicted if the image class was Buy, Hold or Sell, hence a corresponding Algo-trading model was developed.

\begingroup
\footnotesize
\fontsize{7}{9}\selectfont
\begin{longtable}{
                p{0.03\linewidth}
                p{0.18\linewidth}
                p{0.08\linewidth}
                p{0.14\linewidth}
                p{0.14\linewidth}
                p{0.13\linewidth}
                p{0.14\linewidth}
                }
                
\caption{Classification (Buy-sell Signal, or Trend Detection) Based Algo-trading Models}\\

\hline
\textbf{Art.}                                                          & 
\textbf{Data Set}                                                       & 
\textbf{Period}                                                    & 
\textbf{Feature Set}                                                    & 
\textbf{Method}                                                 & 
\textbf{Performance Criteria}                                           & 
\textbf{Environment}                                                      \\ 
\hline
\endhead 

\\

\cite{Sezer_2017} & Stocks in Dow30 & 1997-2017 & \acrshort{rsi}  & \acrshort{dmlp} with genetic algorithm & Annualized return & Spark MLlib, Java \\ \hline
\cite{Navon_2017} & \acrshort{spy} \acrshort{etf}, 10 stocks from \acrshort{sp500} & 2014-2016 & Price data & \acrshort{ffnn} & Cumulative gain & MatConvNet, Matlab \\ \hline
\cite{Troiano_2018} & Dow30 stocks & 2012-2016 & Close data and several technical indicators & \acrshort{lstm} & Accuracy & Python, Keras, Tensorflow, \acrshort{talib} \\ \hline
\cite{Sirignano_2018} & High-frequency record of all orders & 2014-2017 & Price data, record of all orders, transactions & \acrshort{lstm} & Accuracy & - \\ \hline
\cite{Tsantekidis_2017} & Nasdaq Nordic (Kesko Oyj, Outokumpu Oyj, Sampo, Rautaruukki, Wartsila Oyj) & 2010 & Price and volume data in  \acrshort{lob} & \acrshort{lstm} & Precision, Recall, F1-score, Cohen's k & - \\ \hline
\cite{Gudelek_2017} & 17 \acrshortpl{etf} & 2000-2016 & Price data, technical indicators & \acrshort{cnn} & Accuracy, \acrshort{mse}, Profit, \acrshort{auroc} & Keras, Tensorflow \\ \hline
\cite{Sezer_2018} & Stocks in Dow30 and 9 Top Volume \acrshortpl{etf} & 1997-2017 & Price data, technical indicators & \acrshort{cnn} with feature imaging & Recall, precision, F1-score, annualized return & Python, Keras, Tensorflow, Java \\ \hline
\cite{Hu_2018_a} & \acrshort{ftse}100 & 2000-2017 & Price data & \acrshort{cae} & \acrshort{tr}, \acrshort{sr}, \acrshort{mdd}, mean return & - \\ \hline
\cite{Tsantekidis_2017_a} & Nasdaq Nordic (Kesko Oyj, Outokumpu Oyj, Sampo, Rautaruukki, Wartsila Oyj) & 2010  & Price, Volume data, 10 orders of the \acrshort{lob}  & \acrshort{cnn} & Precision, Recall, F1-score, Cohen's k & Theano, Scikit learn, Python \\ \hline

\cite{Gunduz_2017} & Borsa Istanbul 100 Stocks & 2011-2015 & 75 technical indicators and \acrshort{ochlv} & \acrshort{cnn} & Accuracy & Keras \\ \hline
\cite{Sezer_2019} & \acrshortpl{etf} and Dow30 & 1997-2007 & Price data & \acrshort{cnn} with feature imaging & Annualized return & Keras, Tensorflow \\ \hline
\cite{Serrano_2018} & 8 experimental assets from bond/derivative market & - & Asset prices data & \acrshort{rl}, \acrshort{dnn}, Genetic Algorithm & Learning and genetic algorithm error & - \\ \hline
\cite{Saad_1998} & 10 stocks from \acrshort{sp500} & - & Stock Prices & \acrshort{tdnn}, \acrshort{rnn}, \acrshort{pnn} & Missed opportunities, false alarms ratio & - \\ \hline
\cite{Doering_2017} & London Stock Exchange & 2007-2008 & Limit order book state, trades, buy/sell orders, order deletions & \acrshort{cnn} & Accuracy, kappa & Caffe \\ \hline
\cite{Jiang2017deep} & Cryptocurrencies, Bitcoin & 2014-2017 & Price data & \acrshort{cnn}, \acrshort{rnn}, \acrshort{lstm} & Accumulative portfolio value, \acrshort{mdd}, \acrshort{sr} & - \\ \hline

\label{table:algorithmic_trading_2}
\end{longtable}
\endgroup

Serrano et al. \cite{Serrano_2018} proposed a novel method called “GoldAI Sachs” Asset Banker Reinforcement Learning Algorithm for algorithmic trading. The proposed method used a random neural network, \gls{gp}, and \gls{rl} to generate the trading signals.  Saad et al. \cite{Saad_1998} compared  \gls{tdnn}, \gls{rnn} and \gls{pnn} for trend detection using 10 stocks from S\&P500. In  \cite{Doering_2017}, \gls{hft} microstructures forecasting with \gls{cnn} method was performed. In  \cite{Jiang2017deep}, cryptocurrency portfolio management based on three different proposed models (basic \gls{rnn}, \gls{lstm} and \gls{cnn}) was implemented. 

Tino et al. \cite{Tino_2001} used \gls{dax}, \gls{ftse}100, call and put options prices to predict the changes with Markov models and used the financial time series data to predict volatility changes with \gls{rnn}.  Meanwhile, Chen et al. \cite{Chen_2018_c} proposed a method that uses a filterbank \gls{cnn} Algorithm on 15x15 volatility times series converted synthetic images. In the study, the financial domain knowledge and filterbank mechanism were combined to determine the trading signals. Bari et al. \cite{bari_2018} used text mining to extract information from the tweets and financial news and used \gls{lstm}, \gls{rnn}, \gls{gru} for the generation of the trading signals. Dixon et al. \cite{Dixon_2017} used \gls{rnn} for the sequence classification of the limit order book to predict a next event price-flip. 

\begingroup
\footnotesize
\fontsize{7}{9}\selectfont
\begin{longtable}{
                p{0.03\linewidth}
                p{0.18\linewidth}
                p{0.08\linewidth}
                p{0.14\linewidth}
                p{0.14\linewidth}
                p{0.13\linewidth}
                p{0.14\linewidth}
                }
                
\caption{Stand-alone and/or Other Algorithmic Models}\\

\hline
\textbf{Art.}                                                          & 
\textbf{Data Set}                                                       & 
\textbf{Period}                                                    & 
\textbf{Feature Set}                                                    & 
\textbf{Method}                                                 & 
\textbf{Performance Criteria}                                           & 
\textbf{Environment}                                                      \\ 
\hline
\endhead 

\\

\cite{Tino_2001} & \acrshort{dax}, \acrshort{ftse}100, call/put options  & 1991-1998 & Price data & Markov model, \acrshort{rnn}  & Ewa-measure, iv, daily profits' mean and std & - \\ \hline
\cite{Chen_2018_c} & Taiwan Stock Index Futures, Mini Index Futures & 2012-2014 & Price data to image & Visualization method + \acrshort{cnn} & Accumulated profits,accuracy & - \\ \hline
\cite{bari_2018} & Energy-Sector/ Company-Centric Tweets in \acrshort{sp500}  & 2015-2016 & Text and Price data & \acrshort{lstm}, \acrshort{rnn}, \acrshort{gru} & Return, \acrshort{sr}, precision, recall, accuracy & Python,  Tweepy API \\ \hline
\cite{Dixon_2017} & \acrshort{cme} FIX message & 2016 & Limit order book, time-stamp, price data & \acrshort{rnn} & Precision, recall, F1-measure & Python, TensorFlow, R \\ \hline
\cite{Chen_2018_b} & Taiwan stock index futures (TAIFEX) & 2017 & Price data & Agent based \acrshort{rl} with \acrshort{cnn} pre-trained & Accuracy  & - \\ \hline
\cite{Wang_2018_b} & Stocks from \acrshort{sp500} & 2010-2016 & \acrshort{ochlv} & \acrshort{dcnl} & \acrshort{pcc}, \acrshort{dtw}, \acrshort{vwl} & Pytorch \\ \hline
\cite{Day_2016} & News from NowNews, AppleDaily, LTN, MoneyDJ for 18 stocks & 2013-2014 & Text, Sentiment & \acrshort{dnn} & Return & Python, Tensorflow \\ \hline
\cite{Sirignano_2016} & 489 stocks from \acrshort{sp500} and \acrshort{nasdaq}-100 & 2014-2015 & Limit Order Book  & Spatial neural network & Cross entropy error & NVIDIA’s cuDNN \\ \hline
\cite{Gao_2018} & Experimental dataset & - & Price data & \acrshort{drl} with \acrshort{cnn}, \acrshort{lstm}, \acrshort{gru}, \acrshort{mlp} & Mean profit & Python \\ \hline

\label{table:algorithmic_trading_3}
\end{longtable}
\endgroup

Chen et al. \cite{Chen_2018_b} used 1-dimensional \gls{cnn} with an agent-based \gls{rl} algorithm on the Taiwan stock index futures (TAIFEX) dataset. 
Wang et al. \cite{Wang_2018_b} proposed a Deep Co-investment Network Learning (DeepCNL) method that used convolutional and \gls{rnn} layers. 
The investment pattern was determined using the extracted Rise-Fall trends. Day et al. \cite{Day_2016} used financial sentiment analysis using text mining and \gls{dnn} for stock algorithmic trading. Sirignano et al.
  \cite{Sirignano_2016} proposed a ``spatial neural network” model that used limit order book and spatial features for algorithmic trading. Their model estimates the best bid-ask prices using bid, ask prices in the limit order book. Gao et al.
   \cite{Gao_2018} used \gls{gru}, \gls{lstm} units, \gls{cnn}, and \gls{mlp}  to model Q values for the implementation of the \gls{drl} method.

\subsection{Risk Assessment}

Another study area that has been of interest to \gls{dl} researchers is Risk Assessment which identifies the ``riskiness" of any given asset, firm, person, product, bank, etc. Several different versions of this general problem exist, such as bankruptcy prediction, credit scoring, credit evaluation, loan/insurance underwriting, bond rating, loan application, consumer credit determination, corporate credit rating, mortgage choice decision, financial distress prediction, business failure prediction. Correctly identifying the risk status in such cases is crucial, since asset pricing is highly dependent on these risk assessment measures. The mortgage crisis based on improper risk assessment of \gls{cds} between financial institutions caused the real-estate bubble to burst in 2008 and resulted in the Great Recession \cite{Baily_2008}. 

The majority of the risk assessment studies concentrate on credit scoring and bank distress classification. However, there are also a few papers covering mortgage default possibility, risky transaction detection or crisis forecasting. Meanwhile, there are some anomaly detection studies for risk assessment, most of which also fall under the "Fraud Detection" category which will be covered in the next subsection.

\begingroup
\footnotesize
\fontsize{7}{9}\selectfont
\begin{longtable}{
                p{0.03\linewidth}
                p{0.18\linewidth}
                p{0.08\linewidth}
                p{0.14\linewidth}
                p{0.14\linewidth}
                p{0.13\linewidth}
                p{0.14\linewidth}
                }

\caption{Credit Scoring or Classification Studies}\\
\\

\hline
\textbf{Art.}                                                          & 
\textbf{Data Set}                                                       & 
\textbf{Period}                                                    & 
\textbf{Feature Set}                                                    & 
\textbf{Method}                                                 & 
\textbf{Performance Criteria}                                           & 
\textbf{Env.}                                                      \\ 
\hline
\endhead 

\cite{Luo_2017} & The XR 14 \acrshort{cds} contracts & 2016 & Recovery rate, spreads, sector and region & \acrshort{dbn}+\acrshort{rbm} & \acrshort{auroc}, \acrshort{fn}, \acrshort{fp}, Accuracy & WEKA \\ \hline
\cite{Yu_2018} & German, Japanese credit datasets & - & Personal financial variables & \acrshort{svm} + \acrshort{dbn} & Weighted-accuracy, \acrshort{tp}, \acrshort{tn} & - \\ \hline
\cite{Li_2017b} & Credit data from Kaggle & - & Personal financial variables & \acrshort{dnn} & Accuracy, \acrshort{tp}, \acrshort{tn}, G-mean & - \\ \hline
\cite{Tran_2016} & Australian, German credit data & - & Personal financial variables & \acrshort{gp} + \acrshort{ae} as Boosted \acrshort{dnn} & \acrshort{fp} & Python, Scikit-learn \\ \hline
\cite{Neagoe_2018} & German, Australian credit dataset & - & Personal financial variables & \acrshort{dcnn}, \acrshort{mlp} & Accuracy, False/Missed alarm & - \\ \hline
\cite{Zhu_2018} & Consumer credit data from Chinese finance company & - & Relief algorithm chose the 50 most important features & \acrshort{cnn} + Relief & \acrshort{auroc}, K-s statistic, Accuracy & Keras \\ \hline
\cite{Niimi_2015} & Credit approval dataset by UCI Machine Learning repo & - & UCI credit approval dataset & Rectifier, Tanh, Maxout \acrshort{dl} & - & AWS EC2, H2O, R \\ \hline

\label{table:risk_assesment_1}
\end{longtable}
\endgroup

Before going into the details about specific \gls{dl} implementations, it is worthwhile to mention the existing \gls{ml} surveys on the topic. Kirkos et al. \cite{Kirkos_2004}, Ravi et al. \cite{Ravi_2008}, Fethi et al. \cite{Fethi_2010} reviewed the bank performance assessment studies based on \gls{ai} and \gls{ml} models. Lahsasna et al. \cite{Lahsasna_2010}, Chen et al.\cite{Chen_2015s} surveyed the credit scoring and credit risk assessment studies based on soft computing techniques whereas Marques et. al. \cite{Marqus_2013} focused only on \gls{ec} Models for credit scoring implementations. Meanwhile, Kumar et al. \cite{Ravi_Kumar_2007}, Verikas et al. \cite{Verikas_2009} reviewed \gls{ml} implementations of bankruptcy prediction studies. Similarly, Sun et al. \cite{Sun_2014} provided a comprehensive survey about research on financial distress and corporate failures. Apart from these reviews, for assessing overall risk, Lin et al. \cite{Lin_2012} surveyed the financial crisis prediction studies based on \gls{ml} models.

Since risk assessment is becoming vital for survival in today's financial world, a lot of researchers turned their attention to \gls{dl} for higher accuracy. Table~\ref{table:risk_assesment_1}, Table~\ref{table:risk_assesment_2} provide snapshot information about the different risk assessment studies implemented using various \gls{dl} models.

For credit score classification (Table~\ref{table:risk_assesment_1}), Luo et al. \cite{Luo_2017}, used \gls{cds} data for Corporate Credit rating and corresponding credit classification (A,B or C). Among the tested models, \gls{dbn} with \gls{rbm} performed the best. This implementation was probably the first study to implement Credit rating with \gls{dbn}.  Similarly, in \cite{Yu_2018}, a cascaded hybrid model of \gls{dbn}, Backpropagation and \gls{svm} for credit classification was implemented and good performance results (the accuracy was above 80-90 \%) were achieved. In \cite{Li_2017b}, credit risk classification was achieved by using an ensemble of deep \gls{mlp} networks each using subspaces of the whole space by k-means (using minority class in each, but only a partial subspace of the majority class). The data imbalance problem was handled by using multiple subspaces for each classifier, where each of them had all the positive (minor) instances, but a subsample of negative (majority) instances, finally they used an ensemble \hl{of} deep \glspl{mlp} combining each subspace model. In \cite{Tran_2016}, credit scoring was performed using a \gls{sae} network and \gls{gp} model to create credit assessment rules in order to generate good or bad credit cases. In another study, Neagoe et. al. \cite{Neagoe_2018} classified credit scores using various \gls{dmlp} and deep \gls{cnn} networks. In a different study \cite{Zhu_2018}, consumer credit scoring classification was implemented with a 2-D representation of the input consumer data through transforming the data into a 2-D pixel matrix. Then the resulting images were used as the training and test data for \gls{cnn}. 2-D pixel matrix representation of the consumer data was adapted by using \gls{cnn} for image classification. This was the first implementation of credit scoring using \gls{cnn}. Niimi \cite{Niimi_2015} used UCI credit approval dataset \footnote{https://archive.ics.uci.edu/ml/datasets.html} to compare \gls{dl}, \gls{svm}, \gls{lr}, \gls{rf}, \gls{xgboost} and provided information about credit fraud and credit approval applications; then experimented with the credit approval problem with several models. Various models were compared for credit approval classification. Also, some introduction about credit fraud detection was provided.

Financial distress prediction for banks and corporates are studied extensively (Table~\ref{table:risk_assesment_2}). In \cite{Lanbouri_2015}, a hybrid \gls{dbn} with \gls{svm} was used for financial distress prediction to identify whether the firm was in trouble or not, whereas bank risk classification was studied in \cite{Rawte_2018}. In \cite{Ronnqvist_2015}, news semantics were extracted by the word sequence learning and associated events were labeled with the bank stress, then from the formed semantic vector representation, the bank stress was determined and classified against a threshold. Prediction and semantic meaning extraction were integrated in a neat way. In another study \cite{R_nnqvist_2017}, text mining was again used for identifying the bank distress by extracting the data from financial news and then using a \gls{dffn} on semantic sentence vectors extracted from word embeddings to classify if there was an event or not. Similarly, Cerchiello et al. \cite{Cerchiello_2017} used text mining from the financial news to classify bank distress. Malik et al. \cite{malik_2018} evaluated the bank stress by first predicting the bank's performance through an \gls{lstm} network, then Backpropagation network was used for finding the bank stress level. 

\begingroup
\footnotesize
\fontsize{7}{9}\selectfont
\begin{longtable}{
                p{0.03\linewidth}
                p{0.18\linewidth}
                p{0.08\linewidth}
                p{0.14\linewidth}
                p{0.14\linewidth}
                p{0.13\linewidth}
                p{0.14\linewidth}
                }

\caption{Financial Distress, Bankruptcy, Bank Risk, Mortgage Risk, Crisis Forecasting Studies}\\
\\

\hline
\textbf{Art.}                                                          & 
\textbf{Data Set}                                                       & 
\textbf{Period}                                                    & 
\textbf{Feature Set}                                                    & 
\textbf{Method}                                                 & 
\textbf{Performance Criteria}                                           & 
\textbf{Env.}                                                      \\ 
\hline
\endhead 

\cite{Lanbouri_2015} & 966 french firms & - & Financial ratios & \acrshort{rbm}+\acrshort{svm} & Precision, Recall & - \\ \hline
\cite{Rawte_2018} & 883 \acrshort{bhc} from EDGAR & 2006-2017 & Tokens, weighted sentiment polarity, leverage and \acrshort{roa}  & \acrshort{cnn}, \acrshort{lstm}, \acrshort{svm}, \acrshort{rf} & Accuracy, Precision, Recall, F1-score & Keras, Python, Scikit-learn \\ \hline
\cite{Ronnqvist_2015} & The event data set for large European banks, news articles from Reuters & 2007-2014 & Word, sentence & \acrshort{dnn} +\acrshort{nlp} preprocess & Relative usefulness, F1-score & - \\ \hline
\cite{R_nnqvist_2017} & Event dataset on European banks, news from Reuters & 2007-2014 & Text, sentence & Sentence vector + \acrshort{dffn} & Usefulness, F1-score, \acrshort{auroc} & - \\ \hline
\cite{Cerchiello_2017} & News from Reuters, fundamental data & 2007-2014 & Financial ratios and news text & doc2vec + \acrshort{nn} & Relative usefulness & Doc2vec \\ \hline
\cite{malik_2018} & Macro/Micro economic variables, Bank characteristics/performance variables from \acrshort{bhc} & 1976-2017 & Macro economic variables and bank performances & \acrshort{cgan}, \acrshort{mvn}, \acrshort{mv-t}, \acrshort{lstm}, \acrshort{var}, \acrshort{fe-qar} & \acrshort{rmse}, Log likelihood, Loan loss rate & - \\ \hline
\cite{Ribeiro_2011} & Financial statements of French companies  & 2002-2006 & Financial ratios  & \acrshort{dbn} & Recall, Precision, F1-score, \acrshort{fp}, \acrshort{fn} & - \\ \hline
\cite{Yeh_2015} & Stock returns of American publicly-traded companies from \acrshort{crsp} & 2001-2011 & Price data & \acrshort{dbn} & Accuracy & Python, Theano \\ \hline
\cite{Hosaka_2018} & Financial statements of several companies from Japanese stock market & 2002-2016 & Financial ratios & \acrshort{cnn} & F1-score, \acrshort{auroc} & - \\ \hline
\cite{Sirignano_2018_a} & Mortgage dataset with local and national economic factors & 1995-2014 & Mortgage related features & \acrshort{ann} & Negative average log-likelihood & AWS \\ \hline
\cite{Kvamme_2018} & Mortgage data from Norwegian financial service group, DNB & 2012-2016 & Personal financial variables & \acrshort{cnn} & Accuracy, Sensitivity, Specificity, \acrshort{auroc} & - \\ \hline
\cite{Abroyan_2017} & Private brokerage company’s real data of risky transactions & - & 250 features: order details, etc. & \acrshort{cnn}, \acrshort{lstm} & F1-Score & Keras, Tensorflow \\ \hline
\cite{Chatzis_2018} & Several datasets combined to create a new one & 1996-2017 & Index data, 10-year Bond yield, exchange rates, & Logit, \acrshort{cart}, \acrshort{rf}, \acrshort{svm}, \acrshort{nn}, \acrshort{xgboost}, \acrshort{dnn} & \acrshort{auroc}, \acrshort{ks}, \acrshort{g-mean}, likelihood ratio, \acrshort{dpower}, \acrshort{ba}, \acrshort{wba} & R \\ \hline

\label{table:risk_assesment_2}
\end{longtable}
\endgroup

There are also a number of research papers that were focused on bankruptcy or corporate default prediction. Ribeiro et al. \cite{Ribeiro_2011} implemented bankruptcy prediction with \gls{dbn}. The results of \gls{dbn} were compared with \gls{svm} and \gls{rbm}. Yeh et al. \cite{Yeh_2015} used the stock returns of default and solvent companies as inputs to \gls{rbm} used as \gls{sae}, then the output of \gls{rbm} was used as input to \gls{dbn} to predict if the company was solvent or default. The results were compared with an \gls{svm} model and the \gls{dbn} model outperformed \gls{svm}. Hosaka et al. \cite{Hosaka_2018} tried a different approach by converting the financial data to the image to use \gls{cnn} for bankruptcy prediction. 

The remaining implementations of risk assessment are as follows: Sirignano et al. \cite{Sirignano_2018_a} used the mortgage application data of 20 years for identifying the mortgage risk using various parameters. They also performed a lot of analyses relating different factors that affected the mortgage payment structure. The authors also analyzed the prepayment and delinquency behavior in their assessment. For another mortgage risk assessment application, Kvamme et al. \cite{Kvamme_2018} used \gls{cnn} and \gls{rf} models to predict whether a customer would default on its mortgage or not. In a different study, Abroyan et al. \cite{Abroyan_2017} used \gls{cnn} and \gls{lstm} networks to classify if a transaction performed on the stock market (trade) was risky or not and high accuracy was achieved. Finally, Chatzis et al. \cite{Chatzis_2018} developed several \gls{ml} and \gls{dl} models for detecting events that caused the stock market to crash. \gls{dl} models had good classification (detecting crisis or not) performance.

\subsection{Fraud Detection}

Financial fraud is one of the areas where the governments and authorities are desperately trying to find a permanent solution. Several different financial fraud cases exist such as credit card fraud, money laundering, consumer credit fraud, tax evasion, bank fraud, insurance claim fraud. This is one of the most extensively studied areas of finance for \gls{ml} research and several survey papers were published accordingly. At different times, Kirkos et al. \cite{KIRKOS_2007}, Yue et al. \cite{Yue_2007}, Wang et al. \cite{Wang_2010}, Phua et al. \cite{Phua_2010}, Ngai et al. \cite{Ngai_2011}, Sharma et al. \cite{Sharma_2012}, West et al. \cite{West_2016} all reviewed the accounting and financial fraud detection studies based on soft computing and data mining techniques. 

These type of studies mostly can be considered as anomaly detection and are generally classification problems. Table~\ref{table:fraud_detection} presents different fraud detection studies based on \gls{dl} models.   

There are a number of studies focused on identifying credit card fraud. Heryadi et al. \cite{Heryadi_2017} developed several \gls{dl} models for credit card fraud detection for Indonesian banks. They also analyzed the effects of the data imbalance between fraud and nonfraud data. In more recent studies, Roy et al. \cite{Roy_2018} used \gls{lstm} model for the credit card fraud detection, whereas in \cite{Ander_2018}, the authors implemented \gls{mlp} networks to classify if a credit card transaction was fraudulent or not. Sohony et al. \cite{Sohony_2018} used an ensemble of \gls{ffnn} for the detection of card fraud.  Jurgovsky et al. \cite{Jurgovsky_2018} used \gls{lstm} for detecting credit card fraud from credit card transaction sequences. They compared their results with \gls{rf}. 
 
Paula et al. \cite{Paula_2016} used deep \gls{ae} to implement anomaly detection to identify the financial fraud and money laundering for Brazilian companies on export tax claims. In a similar study, Gomes et al. \cite{Gomes_2017} proposed an anomaly detection model that identified the anomalies in parliamentary expenditure spending in Brazilian elections using also deep \gls{ae}. 

Wang et al. \cite{Wang_2018} used text mining and \gls{dnn} models for the detection of automobile insurance fraud. Longfei et al. \cite{Longfei_2017} developed \gls{dnn} models to detect online payment transaction fraud. Costa et al. \cite{Costa_2016}  used character sequences in financial transactions and the responses from the other side to detect if the transaction was fraud or not with \gls{lstm}. Goumagias et al. \cite{Goumagias_2018} used deep Q-learning (\gls{rl}) to predict the risk-averse firms' tax evasion behaviours. Finally, they provided suggestions for the states to maximize their tax revenues accordingly.

\begingroup
\footnotesize
\fontsize{7}{9}\selectfont
\begin{longtable}{
                p{0.03\linewidth}
                p{0.25\linewidth}
                p{0.08\linewidth}
                p{0.20\linewidth}
                p{0.10\linewidth}
                p{0.10\linewidth}
                p{0.07\linewidth}
                }

\caption{Fraud Detection Studies}\\
\\

\hline
\textbf{Art.}                                                          & 
\textbf{Data Set}                                                       & 
\textbf{Period}                                                    & 
\textbf{Feature Set}                                                    & 
\textbf{Method}                                                 & 
\textbf{Performance Criteria}                                           & 
\textbf{Env.}                                                      \\ 
\hline
\endhead 

\cite{Heryadi_2017} & Debit card transactions by a local Indonesia bank & 2016-2017 & Financial transaction amount on several time periods & \acrshort{cnn}, Stacked-\acrshort{lstm}, \acrshort{cnn}-\acrshort{lstm} & \acrshort{auroc} & - \\ \hline
\cite{Roy_2018} & Credit card transactions from retail banking & 2017 & Transaction variables and several derived features & \acrshort{lstm}, \acrshort{gru} & Accuracy & Keras \\ \hline
\cite{Ander_2018} & Card purchases' transactions & 2014-2015 & Probability of fraud per currency/origin country, other fraud related features & \acrshort{ann} & \acrshort{auroc} & - \\ \hline
\cite{Sohony_2018} & Transactions made with credit cards by European cardholders & 2013 & Personal financial variables to \acrshort{pca} & \acrshort{ann}, \acrshort{rf} & Recall, Precision, Accuracy & - \\ \hline
\cite{Jurgovsky_2018} & Credit-card transactions & 2015 & Transaction and bank features & \acrshort{lstm} & \acrshort{auroc} & Keras, Scikit-learn \\ \hline
\cite{Paula_2016} & Databases of foreign trade of the Secretariat of Federal Revenue of Brazil  & 2014 & 8 Features: Foreign Trade, Tax, Transactions, Employees, Invoices, etc & \acrshort{ae} & \acrshort{mse} & H2O, R \\ \hline
\cite{Gomes_2017} & Chamber of Deputies open data, Companies data from Secretariat of Federal Revenue of Brazil & 2009-2017 & 21 features: Brazilian State expense, party name, Type of expense, etc. & Deep Autoencoders & \acrshort{mse}, \acrshort{rmse} & H2O, R \\ \hline
\cite{Wang_2018} & Real-world data for automobile insurance company labeled as fradulent & - & Car, insurance and accident related features & \acrshort{dnn} + \acrshort{lda} & \acrshort{tp}, \acrshort{fp}, Accuracy, Precision, F1-score & - \\ \hline
\cite{Longfei_2017} & Transactions from a giant online payment platform & 2006 & Personal financial variables & \acrshort{gbdt}+\acrshort{dnn} & \acrshort{auroc} & - \\ \hline
\cite{Costa_2016} & Financial transactions & - & Transaction data & \acrshort{lstm} & t-SNE & - \\ \hline
\cite{Goumagias_2018} & Empirical data from Greek firms & - & - & \acrshort{dql} & Revenue & Torch \\ \hline

\label{table:fraud_detection}
\end{longtable}
\endgroup

\subsection{Portfolio Management}

Portfolio Management is the process of choosing various assets within the portfolio for a predetermined period. As seen in other financial applications, slightly different versions of this problem exist, even though the underlying motivation is the same. In general, Portfolio Management covers the following closely related areas: Portfolio Optimization, Portfolio Selection, Portfolio Allocation. Sometimes, these terms are used interchangeably. Li et al. \cite{Li_2014s} reviewed the online portfolio selection studies using various rule-based or \gls{ml} models.

Portfolio Management is actually an optimization problem, identifying the best possible course-of-action for selecting the best-performing assets for a given period. As a result, there are a lot of \gls{ea} models that were developed for this purpose. Metaxiotis et al. \cite{Metaxiotis_2012} surveyed the \glspl{moea} implemented solely on the portfolio optimization problem.

However, some \gls{dl} researchers managed to configure it as a learning model and obtained superior performances. Since Robo-advisory for portfolio management is on the rise, these \gls{dl} implementations have the potential to have a far greater impact on the financial industry in the near future. Table~\ref{table:portfolio_management} presents the portfolio management \gls{dl} models and summarizes their achievements.

There are a number of stock selection implementations. Takeuchi et al. \cite{Takeuchi_2013} classified the stocks in two classes, low momentum and high momentum depending on their expected return. They used a deep \gls{rbm} encoder-classifier network and achieved high returns. Similarly, in \cite{grace_2017}, stocks were evaluated against their benchmark index to classify if they would outperform or underperform using \gls{dmlp}, then based on the predictions, adjusted the portfolio allocation weights for the stocks for enhanced indexing. In \cite{Fu_2018}, an \gls{ml} framework including \gls{dmlp} was constructed and the stock selection problem was implemented. 

\begingroup
\footnotesize
\fontsize{7}{9}\selectfont
\begin{longtable}{
                p{0.03\linewidth}
                p{0.18\linewidth}
                p{0.08\linewidth}
                p{0.14\linewidth}
                p{0.14\linewidth}
                p{0.13\linewidth}
                p{0.14\linewidth}
                }

\caption{Portfolio Management Studies}\\
\\

\hline
\textbf{Art.}                                                          & 
\textbf{Data Set}                                                       & 
\textbf{Period}                                                    & 
\textbf{Feature Set}                                                    & 
\textbf{Method}                                                 & 
\textbf{Performance Criteria}                                           & 
\textbf{Env.}                                                      \\ 
\hline
\endhead

\cite{Jiang2017deep} & Cryptocurrencies, Bitcoin & 2014-2017 & Price data & \acrshort{cnn}, \acrshort{rnn}, \acrshort{lstm} & Accumulative portfolio value, \acrshort{mdd}, \acrshort{sr} & - \\ \hline
\cite{Takeuchi_2013} & Stocks from \acrshort{nyse}, \acrshort{amex}, \acrshort{nasdaq} & 1965-2009 & Price data & Autoencoder + \acrshort{rbm} & Accuracy, confusion matrix & - \\ \hline
\cite{grace_2017} & 20 stocks from \acrshort{sp500} & 2012-2015 & Technical indicators & \acrshort{mlp} & Accuracy & Python, Scikit Learn, Keras, Theano \\ \hline
\cite{Fu_2018} & Chinese stock data & 2012-2013 & Technical, fundamental data & Logistic Regression, \acrshort{rf}, \acrshort{dnn} & \acrshort{auc}, accuracy, precision, recall, f1, tpr, fpr & Keras, Tensorflow, Python, Scikit learn \\ \hline
\cite{Aggarwal_2017} & Top 5 companies in \acrshort{sp500} & - & Price data and Financial ratios  & \acrshort{lstm}, Auto-encoding, Smart indexing & \acrshort{cagr} & - \\ \hline
\cite{Heaton_2016_a} & \acrshort{ibb} biotechnology index,  stocks & 2012-2016 & Price data & Auto-encoding, Calibrating, Validating, Verifying & Returns & - \\ \hline
\cite{Lin_2006} & Taiwans stock market & - & Price data & Elman \acrshort{rnn} & \acrshort{mse}, return & - \\ \hline
\cite{Maknickien__2014} & FOREX (EUR/USD, etc), Gold & 2013 & Price data & Evolino \acrshort{rnn} & Return  & Python \\ \hline
\cite{Zhou_2018_a} & Stocks in \acrshort{nyse}, \acrshort{amex}, \acrshort{nasdaq}, \acrshort{taq} intraday trade & 1993-2017 & Price, 15 firm characteristics & \acrshort{lstm}+\acrshort{mlp} & Monthly return, \acrshort{sr} & Python,Keras, Tensorflow in AWS \\ \hline
\cite{Batres_2015} & \acrshort{sp500} & 1985-2006 & monthly and daily log-returns & \acrshort{dbn}+\acrshort{mlp} & Validation, Test Error & Theano, Python, Matlab \\ \hline
\cite{Lee_2018} & 10 stocks in \acrshort{sp500} & 1997-2016 & \acrshort{ochlv}, Price data  & \acrshort{rnn}, \acrshort{lstm}, \acrshort{gru} & Accuracy, Monthly return & Keras, Tensorflow \\ \hline
\cite{Iwasaki_2018} & Analyst reports on the \acrshort{tse} and Osaka Exchange  & 2016-2018 & Text & \acrshort{lstm}, \acrshort{cnn}, \acrshort{bi-lstm} & Accuracy, \acrshort{r-sq} & R, Python, MeCab \\ \hline
\cite{Liang_2018} & Stocks from Chinese/American stock market & 2015-2018 & \acrshort{ochlv}, Fundamental data & \acrshort{ddpg}, \acrshort{ppo} & \acrshort{sr}, \acrshort{mdd} & - \\ \hline
\cite{Chen_2016} & Hedge fund monthly return data & 1996-2015 & Return, \acrshort{sr}, \acrshort{std}, Skewness, Kurtosis, Omega ratio, Fund alpha & \acrshort{dnn} & Sharpe ratio, Annual return, Cum. return & - \\ \hline
\cite{Jiang_2017} & 12 most-volumed cryptocurrency & 2015-2016 & Price data & \acrshort{cnn} + \acrshort{rl} & \acrshort{sr}, portfolio value, \acrshort{mdd} & - \\ \hline

\label{table:portfolio_management}
\end{longtable}
\endgroup

Portfolio selection and smart indexing were the main focuses of \cite{Aggarwal_2017}  and  \cite{Heaton_2016_a} using \gls{ae} and \gls{lstm} networks. Lin et al. \cite{Lin_2006} used the Elman network for optimal portfolio selection by predicting the stock returns for t+1 and then constructing the optimum portfolio according to the returns. Meanwhile, Maknickiene et al. \cite{Maknickiene_2014} used Evolino \gls{rnn} for portfolio selection and return prediction accordingly. The selected portfolio components (stocks) were orthogonal in nature.

In \cite{Zhou_2018_a}, through predicting the next month's return, top to be performed portfolios were constructed and good monthly returns were achieved with \gls{lstm} and \gls{lstm}-\gls{mlp} combined \gls{dl} models. Similarly, Batres et al. \cite{Batres_2015} combined \gls{dbn} and \gls{mlp} for constructing a stock portfolio by predicting each stock's monthly log-return and choosing the only stocks that were expected to perform better than the performance of the median stock. Lee et al. \cite{Lee_2018} compared 3 \gls{rnn} models (S-\gls{rnn}, \gls{lstm}, \gls{gru}) for stock price prediction and then constructed a threshold-based portfolio with selecting the stocks according to the predictions. With a different approach, Iwasaki et al. \cite{Iwasaki_2018} used the analyst reports for sentiment analyses through text mining and word embeddings and used the sentiment features as inputs to \gls{dfnn} model for the stock price prediction. Then different portfolio selections were implemented based on the projected stock returns.

\gls{drl} was selected as the main \gls{dl} model for \cite{Liang_2018}. Liang et al. \cite{Liang_2018} used \gls{drl} for portfolio allocation by adjusting the stocks weights using various \gls{rl} models. Chen et al. \cite{Chen_2016} compared different \gls{ml} models (including \gls{dffn}) for hedge fund return prediction and hedge fund selection. \gls{dl} and \gls{rf} models had the best performance.

Cryptocurrency portfolio management also started getting attention from \gls{dl} researchers. In \cite{Jiang_2017}, portfolio management (allocation and adjustment of weights) was implemented by \gls{cnn} and \gls{drl} on selected cryptocurrencies. Similarly, Jiang et al. \cite{Jiang2017deep} implemented cryptocurrency portfolio management (allocation) based on 3 different proposed models, namely \gls{rnn}, \gls{lstm} and \gls{cnn}.

\subsection{Asset Pricing and Derivatives Market (options, futures, forward contracts)}

Accurate pricing or valuation of an asset is a fundamental study area in finance. There are a vast number of \gls{ml} models developed for banks, corporates, real estate, derivative products, etc. However, \gls{dl} has not been applied to this particular field and there are some possible implementation areas that \gls{dl} models can assist the asset pricing researchers or valuation experts. There were only a handful of studies that we were able to pinpoint within the \gls{dl} and finance community. There are vast opportunities in this field for future studies and publications. 

Meanwhile, financial models based on derivative products is quite common. Options pricing, hedging strategy development,  financial engineering with options, futures, forward contracts are among some of the studies that can benefit from developing \gls{dl} models. Some recent studies indicate that researchers started showing interest in \gls{dl} models that can provide solutions to this complex and challenging field. Table~\ref{table:derivatives_market} summarizes these studies with their intended purposes. 

\begingroup
\footnotesize
\fontsize{7}{9}\selectfont
\begin{longtable}{
                p{0.04\linewidth}
                p{0.08\linewidth}
                p{0.13\linewidth}
                p{0.08\linewidth}
                p{0.15\linewidth}
                p{0.13\linewidth}
                p{0.13\linewidth}
                p{0.08\linewidth}
                }

\caption{Asset Pricing and Derivatives Market Studies}\\
\\

\hline
\textbf{Art.}                                                          & 
\textbf{Der.Type}                                                & 
\textbf{Data Set}                                                       & 
\textbf{Period}                                                    & 
\textbf{Feature Set}                                                    & 
\textbf{Method}                                                 & 
\textbf{Performance Criteria}                                           & 
\textbf{Env.}                                                      \\ 
\hline
\endhead 

\cite{Iwasaki_2018} & Stock exchange & Analyst reports on the \acrshort{tse} and Osaka Exchange  & 2016-2018 & Text & \acrshort{lstm}, \acrshort{cnn}, \acrshort{bi-lstm} & Accuracy, \acrshort{r-sq} & R, Python, MeCab \\ \hline
\cite{Culkin_2017} & Options & Simulated a range of call option prices & - & Price data, option strike/maturity, dividend/risk free rates, volatility & \acrshort{dnn} & \acrshort{rmse}, the average percentage pricing error & Tensorflow \\ \hline
\cite{Hsu_2018} & Futures, Options & \acrshort{taiex} Options & 2017 & \acrshort{ochlv}, fundamental analysis, option price & \acrshort{mlp}, \acrshort{mlp} with Black scholes & \acrshort{rmse}, \acrshort{mae}, \acrshort{mape} & - \\ \hline
\cite{Feng_2018_a} & Equity returns & Returns in \acrshort{nyse}, \acrshort{amex}, \acrshort{nasdaq} & 1975-2017 & 57 firm characteristics & Fama-French n-factor model \acrshort{dl} & \acrshort{r-sq},\acrshort{rmse} & Tensorflow \\ \hline

\label{table:derivatives_market}
\end{longtable}
\endgroup

Iwasaki et al. \cite{Iwasaki_2018} \hl{used a} \gls{dfnn} model and the analyst reports for sentiment analyses to predict the stock prices. Different \hl{portfolio selection approaches} were implemented after the prediction of the stock prices. Culkin et al. \cite{Culkin_2017} proposed a novel method that used feedforward \gls{dnn} model to predict option prices by comparing their results with Black \& Scholes option pricing formula. Similarly, Hsu et al. \cite{Hsu_2018} proposed a novel method that predicted TAIEX option prices using bid-ask spreads and Black \& Scholes option price model parameters with 3-layer \gls{dmlp}. In \cite{Feng_2018_a}, characteristic features such as Asset growth, Industry momentum, Market equity, Market Beta, etc. were used as inputs to a Fama-French n-factor model \gls{dl} to predict US equity returns in \gls{nasdaq}, \gls{amex}, \gls{nyse} indices.

\subsection{Cryptocurrency and Blockchain Studies}

In the last few years, cryptocurrencies have been the talk of the town due to their incredible price gain and loss within short periods. Even though price forecasting dominates the area of interest, some other studies also exist, such as cryptocurrency Algo-trading models. 

Meanwhile, Blockchain is a new technology that provides a distributed decentralized ledger system that fits well with the cryptocurrency world. As a matter of fact, cryptocurrency and blockchain are highly coupled, even though blockchain technology has a much wider span for various implementation possibilities that need to be studied. It is still in its early development phase, hence there is a lot of hype in its potentials. 

Some \gls{dl} models have already appeared about cryptocurrency studies, mostly price prediction or trading systems. However, still there is a lack of studies for blockchain research within the \gls{dl} community. Given the attention that the underlying technology has attracted, there is a great chance that some new studies will start appearing in the near future. Table~\ref{table:cryptocurrency_and_blockchain} tabulates the studies for the cryptocurrency and blockchain research.

\begingroup
\footnotesize
\fontsize{7}{9}\selectfont
\begin{longtable}{
                p{0.03\linewidth}
                p{0.13\linewidth}
                p{0.08\linewidth}
                p{0.15\linewidth}
                p{0.15\linewidth}
                p{0.13\linewidth}
                p{0.10\linewidth}
                }

\caption{Cryptocurrency and Blockchain Studies}\\
\\

\hline
\textbf{Art.}                                                          & 
\textbf{Data Set}                                                       & 
\textbf{Period}                                                    & 
\textbf{Feature Set}                                                    & 
\textbf{Method}                                                 & 
\textbf{Performance Criteria}                                           & 
\textbf{Env.}                                                      \\ 
\hline
\endhead

\cite{Spilak_2018} & Bitcoin, Dash, Ripple, Monero, Litecoin, Dogecoin, Nxt, Namecoin & 2014-2017 & \acrshort{ma}, \acrshort{boll}, the \acrshort{crix} daily returns, Euribor interest rates, \acrshort{ochlv} of EURO/UK, EURO/USD, US/JPY & \acrshort{lstm}, \acrshort{rnn}, \acrshort{mlp} & Accuracy, F1-measure & Python, Tensorflow \\ \hline
\cite{Jiang2017deep} & Cryptocurrencies, Bitcoin & 2014-2017 & Price data &  \acrshort{cnn} & Accumulative portfolio value, \acrshort{mdd}, \acrshort{sr} & - \\ \hline
\cite{Jiang_2017} & 12 most-volumed cryptocurrency & 2015-2016 & Price data & \acrshort{cnn} + \acrshort{rl} & \acrshort{sr}, portfolio value, \acrshort{mdd} &  \\ \hline
\cite{Chen_2018_a} & Bitcoin data & 2010-2017 & Hash value, bitcoin address, public/private key, digital signature, etc. & Takagi–Sugeno Fuzzy cognitive maps  & Analytical hierarchy process & - \\ \hline
\cite{Nan_2018} & Bitcoin data & 2012, 2013, 2016 & TransactionId, input/output Addresses, timestamp & Graph embedding using heuristic, laplacian eigen-map, deep \acrshort{ae} & F1-score & - \\ \hline
\cite{Lopes_2018_thesis} & Bitcoin, Litecoin, StockTwits & 2015-2018 & \acrshort{ochlv}, technical indicators, sentiment analysis &  \acrshort{cnn}, \acrshort{lstm}, State Frequency Model & \acrshort{mse} & Keras, Tensorflow \\ \hline
\cite{McNally_2018} & Bitcoin  & 2013-2016 & Price data & Bayesian optimized \acrshort{rnn}, \acrshort{lstm} & Sensitivity, specificity, precision, accuracy, \acrshort{rmse} & Keras, Python, Hyperas \\ \hline

\label{table:cryptocurrency_and_blockchain}
\end{longtable}
\endgroup

Chen et al. \cite{Chen_2018_a} proposed a blockchain transaction traceability algorithm using Takagi-Sugeno fuzzy cognitive map and 3-layer \gls{dmlp}. Bitcoin data (Hash value, bitcoin address, public/private key, digital signature, etc.) was used as the dataset. Nan et al. \cite{Nan_2018} proposed a method for bitcoin mixing detection that consisted of different stages: Constructing the Bitcoin transaction graph, \hl{implementing node embedding}, \hl{detecting outliers} through \gls{ae}. Lopes et al. \cite{Lopes_2018_thesis} combined the opinion market and price prediction for cryptocurrency trading. Text mining combined with 2 models, \gls{cnn} and \gls{lstm} were used to extract the opinion. Bitcoin, Litecoin, StockTwits were used as the dataset. \gls{ochlv} of prices, technical indicators, and sentiment analysis were used as the feature set.

In another study, Jiang et al. \cite{Jiang2017deep} presented a financial-model-free \gls{rl} framework for the Cryptocurrency portfolio management that was based on 3 different proposed models, basic \gls{rnn}, \gls{lstm} and \gls{cnn}. In \cite{Jiang_2017}, portfolio management was implemented by \gls{cnn} and \gls{drl} on 12 most-volumed cryptocurrencies. Bitcoin, Ethereum, Bitcoin Cash and Digital Cash were used as the dataset. 

In addition, Spilak et al. \cite{Spilak_2018} used 8 cryptocurrencies (Bitcoin, Dash, Ripple, Monero, Litecoin, Dogecoin, Nxt, Namecoin) to construct a dynamic portfolio using \gls{lstm}, \gls{rnn}, \gls{mlp} methods. McNally et al. \cite{McNally_2018} compared Bayesian optimized \gls{rnn}, \gls{lstm} and \gls{arima} to predict the bitcoin price direction. Sensitivity, specificity, precision, accuracy, \gls{rmse} were used as the performance metrics.

\subsection{Financial Sentiment Analysis and Behavioral Finance}

One of the most important components of behavioral finance is emotion or investor sentiment. Lately, advancements in text mining techniques opened up the possibilities for successful sentiment extraction through social media feeds. There is a growing interest in Financial Sentiment Analysis, especially for trend forecasting and Algo-trading model development. Kearney et al. \cite{Kearney_2014} surveyed \gls{ml}-based financial sentiment analysis studies that use textual data.

Nowadays there is broad interest in the sentiment analysis for financial forecasting research using \gls{dl} models. Table~\ref{table:financial_sentiment_analysis} provides information about the sentiment analysis studies that are focused on financial forecasting and based on text mining.

In \cite{Wang_2018_a}, technical analysis (\gls{macd}, \gls{ma}, \gls{dmi}, \gls{ema}, \gls{tema}, Momentum, \gls{rsi}, \gls{cci}, Stochastic Oscillator, \gls{roc}) and sentiment analysis (using social media) were used to predict the price of stocks. Shi et al.  \cite{Shi_2018} proposed a method that visually interpreted text-based \gls{dl} models in predicting the stock price movements. They used the financial news from Reuters and Bloomberg. In \cite{Peng_2016}, text mining and word embeddings were used to extract information from the financial news from Reuters and Bloomberg to predict the stock price movements. In addition, in \cite{Zhuge_2017}, the prices of index data and emotional data from text posts were used to predict the stock opening price of the next day. \hl{Zhongshengz} \cite{Zhongshengz_2018} performed classification and stock price prediction using text and price data. Das et al. \cite{Das_2018} used Twitter sentiment data and stock price data to predict the prices of Google, Microsoft and Apple stocks. 

Prosky et al. \cite{Prosky_2017} performed sentiment, mood prediction using news from Reuters and used these sentiments for price prediction. Li et al. \cite{Jiahong_Li_2017} used sentiment classification (neutral, positive, negative) for the stock open or close price prediction with \gls{lstm} (various models). They compared their results with \gls{svm} and achieved higher overall performance. Iwasaki et al. \cite{Iwasaki_2018} used analyst reports for sentiment analysis through text mining and word embeddings. They used the sentiment features as inputs to \gls{dfnn} model for stock price prediction. Finally, different portfolio selections were implemented based on the projected stock returns.

In a different study, Huang et al. \cite{Huang_2016} used several models including \gls{hmm}, \gls{dmlp} and \gls{cnn} using Twitter moods along with the financial price data for prediction of the next day's move (up or down). \gls{cnn} achieved the best result. 

\begingroup
\footnotesize
\fontsize{7}{9}\selectfont
\begin{longtable}{
                p{0.03\linewidth}
                p{0.20\linewidth}
                p{0.08\linewidth}
                p{0.15\linewidth}
                p{0.10\linewidth}
                p{0.13\linewidth}
                p{0.10\linewidth}
                }

\caption{Financial Sentiment Studies coupled with Text Mining for Forecasting}\\
\\

\hline
\textbf{Art.}                                                          & 
\textbf{Data Set}                                                       & 
\textbf{Period}                                                    & 
\textbf{Feature Set}                                                    & 
\textbf{Method}                                                 & 
\textbf{Performance Criteria}                                           & 
\textbf{Env.}                                                      \\ 
\hline
\endhead

\cite{Iwasaki_2018} & Analyst reports on the \acrshort{tse} and Osaka Exchange  & 2016-2018 & Text & \acrshort{lstm}, \acrshort{cnn}, \acrshort{bi-lstm} & Accuracy, \acrshort{r-sq} & R, Python, MeCab \\ \hline
\cite{Wang_2018_a} & Sina Weibo, Stock market records & 2012-2015 & Technical indicators, sentences & \acrshort{drse} & F1-score, precision, recall, accuracy, \acrshort{auroc} & Python \\ \hline
\cite{Shi_2018} & News from Reuters and Bloomberg for \acrshort{sp500} stocks & 2006-2015 & Financial news, price data & DeepClue & Accuracy & Dynet software \\ \hline
\cite{Peng_2016} & News from Reuters and Bloomberg, Historical stock security data & 2006-2013 & News, price data & \acrshort{dnn} & Accuracy & - \\ \hline
\cite{Zhuge_2017} & \acrshort{sci} prices & 2008-2015 & \acrshort{ochl} of change rate, price & Emotional Analysis + \acrshort{lstm} & \acrshort{mse} & - \\ \hline
\cite{Zhongshengz_2018} & \acrshort{sci} prices & 2013-2016 & Text data and Price data & \acrshort{lstm} & Accuracy, F1-Measure & Python, Keras \\ \hline
\cite{Das_2018} & Stocks of Google, Microsoft and Apple & 2016-2017 & Twitter sentiment and stock prices & \acrshort{rnn} & - & Spark, Flume,Twitter API, \\ \hline
\cite{Prosky_2017} & 30 \acrshort{djia} stocks, \acrshort{sp500}, \acrshort{dji}, news from Reuters & 2002-2016 & Price data and features from news articles & \acrshort{lstm}, \acrshort{nn}, \acrshort{cnn} and word2vec & Accuracy & VADER \\ \hline
\cite{Jiahong_Li_2017} & Stocks  of \acrshort{csi}300 index, \acrshort{ochlv} of \acrshort{csi}300 index & 2009-2014 & Sentiment Posts, Price data & Naive Bayes + \acrshort{lstm} & Precision, Recall, F1-score, Accuracy & Python, Keras \\ \hline
\cite{Huang_2016} & \acrshort{sp500}, \acrshort{nyse} Composite, \acrshort{djia}, \acrshort{nasdaq} Composite & 2009-2011 & Twitter moods, index data & \acrshort{dnn}, \acrshort{cnn} & Error rate  & Keras, Theano \\ \hline

\label{table:financial_sentiment_analysis}
\end{longtable}
\endgroup

Even though financial sentiment is highly coupled with text mining, we decided to represent those two topics in different subsections. The main reason for such a choice is not only the existence of some financial sentiment studies which do not directly depend on financial textual data (like \cite{Huang_2016}) but also the existence of some financial text mining studies that are not automatically used for sentiment analysis which will be covered in the next section.   

\subsection{Financial Text Mining}

With the rapid spreading of social media and real-time streaming news/tweets, instant text-based information retrieval became available for financial model development. As a result, financial text mining studies became very popular in recent years. Even though some of these studies are directly interested in the sentiment analysis through crowdsourcing, there are a lot of implementations that are interested in the content retrieval of news, financial statements, disclosures, etc. through analyzing the text context. There are a few \gls{ml} surveys focused on text mining and news analytics. Among the noteworthy studies of such, Mitra et al. \cite{Mitra_2012} edited a book on news analytics in finance, whereas Li et al. \cite{Li_2011s}, Loughran et al. \cite{Loughran_2016}, Kumar et al. \cite{Kumar_2016} surveyed the studies of textual analysis of financial documents, news and corporate disclosures.  It is worth to mention that there are also some studies  \cite{Mittermayer_2006, Nassirtoussi_2014} of text mining for financial prediction models. 

Previous section was focused on \gls{dl} models using sentiment analysis specifically tailored for the financial forecasting implementations, whereas this section will include \gls{dl} studies that have text Mining without Sentiment Analysis for Forecasting (Table~\ref{table:financial_text_mining_1}), financial sentiment analysis coupled with text mining without forecasting \hl{intent} (Table~\ref{table:financial_text_mining_2}) and finally other text mining implementations (Table~\ref{table:financial_text_mining_3}), respectively. 

Huynh et al. \cite{Huynh_2017} used the financial news from Reuters, Bloomberg and stock prices data to predict the stock movements in the future. In \cite{Han_2018}, different event-types on Chinese companies are classified based on a novel event-type pattern classification algorithm. Besides, the stock prices were predicted using additional inputs. Kraus et al. \cite{Kraus_2017} implemented \gls{lstm} with transfer learning using text mining through financial news and stock market data. Dang et al. \cite{Dang_2018} used Stock2Vec and \gls{tgru} models to generate the input data from the financial news and stock prices for the classification of stock prices. 

In \cite{Ding_2015}, events were detected from Reuters and Bloomberg news through text mining. The extracted information was used for price prediction and stock trading through the \gls{cnn} model. Vargas et al. \cite{Vargas_2017} used text mining and price prediction together for intraday directional movement estimation. Akita et al. \cite{Akita_2016} implemented a method that used text mining and price prediction together for forecasting prices. Verma et al. \cite{Verma_2017} combined news data with financial data to classify the stock price movement. Bari et al. \cite{bari_2018} used text mining for extracting information from the tweets and news. In the method, time series models were used for stock trade signal generation. In  \cite{Zhang_2018}, a method that performed information fusion from news and social media sources was proposed to predict the trend of the stocks. 

In \cite{Chen_2018_e}, social media news were used to predict the index price and the index direction with \gls{rnn}-Boost through \gls{lda} features. Hu et al. \cite{Hu_2018} proposed a novel method that used text mining techniques and Hybrid Attention Networks based on the financial news for forecasting the trend of stocks. Li et al.  \cite{Li_2018} implemented intraday stock price direction classification using the financial news and stocks prices. In \cite{Lee_2017_b}, financial news data and word embedding with Word2vec were implemented to create the inputs for \gls{rcnn} to predict the stock price.

\begingroup
\footnotesize
\fontsize{7}{9}\selectfont
\begin{longtable}{
                p{0.03\linewidth}
                p{0.20\linewidth}
                p{0.08\linewidth}
                p{0.13\linewidth}
                p{0.10\linewidth}
                p{0.13\linewidth}
                p{0.10\linewidth}
                }

\caption{Text Mining Studies without Sentiment Analysis for Forecasting}\\
\\

\hline
\textbf{Art.}                                                          & 
\textbf{Data Set}                                                       & 
\textbf{Period}                                                    & 
\textbf{Feature Set}                                                    & 
\textbf{Method}                                                 & 
\textbf{Performance Criteria}                                           & 
\textbf{Env.}                                                     \\ 
\hline
\endhead

\cite{bari_2018} & Energy-Sector/ Company-Centric Tweets in \acrshort{sp500}  & 2015-2016 & Text and Price data &  & Return, \acrshort{sr}, precision, recall, accuracy & Python,  Tweepy API \\ \hline
\cite{Huynh_2017} & News from Reuters, Bloomberg & 2006-2013 & Financial news, price data & \acrshort{bi-gru} & Accuracy & Python, Keras \\ \hline
\cite{Han_2018} & News from Sina.com, ACE2005 Chinese corpus & 2012-2016 & A set of news text & Their unique algorithm & Precision, Recall, F1-score  & - \\ \hline
\cite{Kraus_2017} & \acrshort{cdax} stock market data & 2010-2013 & Financial news,  stock market data & \acrshort{lstm} & \acrshort{mse}, \acrshort{rmse}, \acrshort{mae}, Accuracy,  \acrshort{auc} & TensorFlow, Theano, Python, Scikit-Learn \\ \hline
\cite{Dang_2018} & Apple, Airbus, Amazon news from Reuters, Bloomberg, \acrshort{sp500} stock prices & 2006-2013 & Price data, news, technical indicators & \acrshort{tgru}, stock2vec & Accuracy, precision, \acrshort{auroc} & Keras, Python \\ \hline
\cite{Ding_2015} & \acrshort{sp500} Index,  15 stocks in \acrshort{sp500} & 2006-2013 & News from Reuters and Bloomberg & \acrshort{cnn} & Accuracy, \acrshort{mcc} & - \\ \hline
\cite{Vargas_2017} & \acrshort{sp500} index news from Reuters & 2006-2013 & Financial news titles, Technical indicators & SI-RCNN (\acrshort{lstm} + \acrshort{cnn}) & Accuracy & - \\ \hline
\cite{Akita_2016} & 10 stocks in Nikkei 225 and news & 2001-2008 & Textual information and Stock prices & Paragraph Vector + \acrshort{lstm} & Profit & - \\ \hline
\cite{Verma_2017} & \acrshort{nifty}50 Index, \acrshort{nifty} Bank/Auto/IT/Energy Index, News & 2013-2017 & Index data, news & \acrshort{lstm} & \acrshort{mcc}, Accuracy & - \\ \hline
\cite{Zhang_2018} & Price data, index data, news, social media data & 2015 & Price data, news from articles and social media & Coupled matrix and tensor & Accuracy, \acrshort{mcc} & Jieba \\ \hline
\cite{Chen_2018_e} & \acrshort{hs}300 & 2015-2017 & Social media news, price data & \acrshort{rnn}-Boost with \acrshort{lda} & Accuracy, \acrshort{mae}, \acrshort{mape}, \acrshort{rmse} & Python, Scikit-learn \\ \hline
\cite{Hu_2018} & News and Chinese stock data & 2014-2017 & Selected words in a news & \acrshort{han} & Accuracy, Annual return & - \\ \hline
\cite{Li_2018} & News, stock prices from Hong Kong Stock Exchange & 2001 & Price data and \acrshort{tfidf} from news & \acrshort{elm}, \acrshort{dlr}, \acrshort{pca}, \acrshort{belm}, \acrshort{kelm}, \acrshort{nn} & Accuracy & Matlab \\ \hline
\cite{Lee_2017_b} & \acrshort{twse} index, 4 stocks in \acrshort{twse} & 2001-2017 & Technical indicators, Price data, News & \acrshort{cnn} + \acrshort{lstm} & \acrshort{rmse}, Profit & Keras, Python, \acrshort{talib} \\ \hline
\cite{Minami_2018} & Stock of Tsugami Corporation & 2013 & Price data & \acrshort{lstm} & \acrshort{rmse} & Keras, Tensorflow \\ \hline
\cite{Yoshihara_2014} & News, Nikkei Stock Average and 10-Nikkei companies & 1999-2008 & news, \acrshort{macd} & \acrshort{rnn}, \acrshort{rbm}+\acrshort{dbn} & Accuracy, P-value & - \\ \hline
\cite{Buczkowski_2017} & ISMIS 2017 Data Mining Competition dataset & - &  Expert identifier, classes & \acrshort{lstm} + \acrshort{gru} + \acrshort{ffnn} & Accuracy & - \\ \hline
\cite{Pinheiro_2017} & Reuters, Bloomberg News, \acrshort{sp500} price & 2006-2013 & News and sentences & \acrshort{lstm} & Accuracy & - \\ \hline
\cite{Liu_2018} & APPL from \acrshort{sp500} and news from Reuters & 2011-2017 & Input news, \acrshort{ochlv}, Technical indicators & \acrshort{cnn} + \acrshort{lstm}, \acrshort{cnn}+\acrshort{svm} & Accuracy, F1-score & Tensorflow \\ \hline
\cite{MATSUBARA_2018} & Nikkei225, \acrshort{sp500}, news from Reuters and Bloomberg & 2001-2013 & Stock price data and news & \acrshort{dgm} & Accuracy, \acrshort{mcc}, \%profit & - \\ \hline
\cite{Nascimento_2015} & Stocks from \acrshort{sp500} & 2006-2013 & Text (news) and Price data & \acrshort{lar}+News, \acrshort{rf}+News & \acrshort{mape}, \acrshort{rmse} & - \\ \hline

\label{table:financial_text_mining_1}
\end{longtable}
\endgroup

Minami et al. \cite{Minami_2018} proposed a method that predicted the stock price with corporate action event information and macro-economic index data using \gls{lstm}. In \cite{Yoshihara_2014}, a novel method that used a combination of \gls{rbm}, \gls{dbn} and word embeddings to create word vectors for \gls{rnn}-\gls{rbm}-\gls{dbn} network was proposed to predict the stock prices. Buczkowski et al. \cite{Buczkowski_2017} proposed a novel method that used expert recommendations, ensemble of \gls{gru} and \gls{lstm} for prediction of the prices.

In \cite{Pinheiro_2017} a novel method that used character-based neural language model using financial news and \gls{lstm} was proposed. Liu et al. \cite{Liu_2018} proposed a method that used word embeddings with word2Vec, technical analysis features and stock prices for price prediction. In  \cite{MATSUBARA_2018}, \gls{dgm} with news articles using Paragraph Vector algorithm was used for creation of the input vector to predict the stock prices.  In \cite{Nascimento_2015}, the stock price data and word embeddings were used for stock price prediction. The results showed that the extracted information from embedding news improves the performance.

\begingroup
\footnotesize
\fontsize{7}{9}\selectfont
\begin{longtable}{
                p{0.03\linewidth}
                p{0.20\linewidth}
                p{0.08\linewidth}
                p{0.13\linewidth}
                p{0.10\linewidth}
                p{0.13\linewidth}
                p{0.10\linewidth}
                }

\caption{Financial Sentiment Studies coupled with Text Mining without Forecasting}\\
\\

\hline
\textbf{Art.}                                                          & 
\textbf{Data Set}                                                       & 
\textbf{Period}                                                    & 
\textbf{Feature Set}                                                    & 
\textbf{Method}                                                 & 
\textbf{Performance Criteria}                                           & 
\textbf{Env.}                                                     \\ 
\hline
\endhead

\cite{Rawte_2018} & 883 \acrshort{bhc} from EDGAR & 2006-2017 & Tokens, weighted sentiment polarity, leverage and \acrshort{roa}  & \acrshort{cnn}, \acrshort{lstm}, \acrshort{svm}, Random Forest & Accuracy, Precision, Recall, F1-score & Keras, Python, Scikit-learn \\ \hline
\cite{Akhtar_2017} & SemEval-2017 dataset, financial text, news, stock market data & 2017 & Sentiments in Tweets, News headlines & Ensemble \acrshort{svr}, \acrshort{cnn}, \acrshort{lstm}, \acrshort{gru} & Cosine similarity score, agreement score, class score & Python, Keras, Scikit Learn \\ \hline
\cite{Chang_2016} & Financial news from Reuters & 2006-2015 & Word vector, Lexical and Contextual input & Targeted dependency tree \acrshort{lstm} & Cumulative abnormal return & - \\ \hline
\cite{Jangid_2018} & Stock sentiment analysis from StockTwits & 2015 & StockTwits messages & \acrshort{lstm}, Doc2Vec, \acrshort{cnn} & Accuracy, precision, recall, f-measure, \acrshort{auc} & - \\ \hline
\cite{Shijia_2018} & Sina Weibo, Stock market records & 2012-2015 & Technical indicators, sentences & \acrshort{drse} & F1-score, precision, recall, accuracy, \acrshort{auroc} & Python \\ \hline
\cite{Sohangir_2018_a} & News from NowNews, AppleDaily, LTN, MoneyDJ for 18 stocks & 2013-2014 & Text, Sentiment &  & Return & Python, Tensorflow \\ \hline
\cite{Mahmoudi_2018} & StockTwits & 2008-2016 & Sentences,  StockTwits messages & \acrshort{cnn}, \acrshort{lstm}, \acrshort{gru} & \acrshort{mcc}, \acrshort{wsurt} & Keras, Tensorflow \\ \hline
\cite{Kitamori_2017} & Financial statements of Japan companies & - & Sentences, text & \acrshort{dnn} & Precision, recall, f-score & - \\ \hline
\cite{Piao_2018} & Twitter posts, news headlines & - & Sentences, text & \acrshort{deep-fasp} & Accuracy, \acrshort{mse}, \acrshort{r-sq} & - \\ \hline
\cite{Li_2014} & Forums data & 2004-2013 & Sentences and keywords & Recursive neural tensor networks & Precision, recall, f-measure & - \\ \hline
\cite{Moore_2017} & News from  Financial Times related US stocks & - & Sentiment of news headlines  & \acrshort{svr}, Bidirectional \acrshort{lstm} & Cosine similarity & Python, Scikit Learn, Keras, Tensorflow \\ \hline

\label{table:financial_text_mining_2}
\end{longtable}
\endgroup

Akhtar et al. \cite{Akhtar_2017} compared \gls{cnn}, \gls{lstm} and \gls{gru} based \gls{dl} models against \gls{mlp} for financial sentiment analysis. Rawte et al. \cite{Rawte_2018} tried to solve three separate problems using \gls{cnn}, \gls{lstm}, \gls{svm}, \gls{rf}: Bank risk classification, sentiment analysis and \gls{roa} regression.

Chang et al. \cite{Chang_2016} implemented the estimation of information content polarity (negative/positive effect) with text mining, word vector, lexical, contextual input and various \gls{lstm} models. They used the financial news from Reuters.  

Jangid et al. \cite{Jangid_2018} proposed a novel method that is a combination of \gls{lstm} and \gls{cnn} for word embedding and sentiment analysis using \gls{bi-lstm} for aspect extraction. The proposed method used multichannel \gls{cnn} for financial sentiment analysis. Shijia et al. \cite{Shijia_2018} used an attention-based \gls{lstm} for the financial sentiment analysis using news headlines and microblog messages. Sohangir et al. \cite{Sohangir_2018_a} used \gls{lstm}, doc2vec, \gls{cnn} and stock market opinions posted in StockTwits for sentiment analysis. Mahmoudi et al. \cite{Mahmoudi_2018} extracted tweets from StockTwits to identify the user sentiment. In the evaluation approach, they also used emojis for the sentiment analysis. Kitamori et al.  \cite{Kitamori_2017} extracted the sentiments from financial news and used \gls{dnn} to classify positive and negative news.

In \cite{Piao_2018}, the sentiment/aspect prediction was implemented using an ensemble of \gls{lstm}, \gls{cnn} and \gls{gru} networks. In a different study, Li et al. \cite{Li_2014}  proposed  a  \gls{dl} based sentiment analysis method using \gls{rnn} to identify the top sellers in the underground economy. Moore et al. \cite{Moore_2017} used text mining techniques for sentiment analysis from the financial news. 

\begingroup
\footnotesize
\fontsize{7}{9}\selectfont
\begin{longtable}{
                p{0.03\linewidth}
                p{0.20\linewidth}
                p{0.08\linewidth}
                p{0.13\linewidth}
                p{0.10\linewidth}
                p{0.13\linewidth}
                p{0.10\linewidth}
                }

\caption{Other Text Mining Studies}\\
\\

\hline
\textbf{Art.}                                                          & 
\textbf{Data Set}                                                       & 
\textbf{Period}                                                    & 
\textbf{Feature Set}                                                    & 
\textbf{Method}                                                 & 
\textbf{Performance Criteria}                                           & 
\textbf{Env.}                                                     \\ 
\hline
\endhead

\cite{Day_2016} & News from NowNews, AppleDaily, LTN, MoneyDJ for 18 stocks & 2013-2014 & Text, Sentiment &  & Return & Python, Tensorflow \\ \hline
\cite{Ronnqvist_2015} & The event data set for large European banks, news articles from Reuters & 2007-2014 & Word, sentence & \acrshort{dnn} +\acrshort{nlp} preprocess & Relative usefulness, F1-score & - \\ \hline
\cite{R_nnqvist_2017} & Event dataset on European banks, news from Reuters & 2007-2014 & Text, sentence & Sentence vector + \acrshort{dffn} & Usefulness, F1-score, \acrshort{auroc} & - \\ \hline
\cite{Cerchiello_2017} & News from Reuters, fundamental data & 2007-2014 & Financial ratios and news text & doc2vec + \acrshort{nn} & Relative usefulness & Doc2vec \\ \hline
\cite{Wang_2018} & Real-world data for automobile insurance company labeled as fradulent & - & Car, insurance and accident related features & \acrshort{dnn} + \acrshort{lda} & \acrshort{tp}, \acrshort{fp}, Accuracy, Precision, F1-score & - \\ \hline
\cite{Costa_2016} & Financial transactions & - & Transaction data & \acrshort{lstm} & t-SNE & - \\ \hline
\cite{Ying_2017} & Taiwan’s National Pension Insurance & 2008-2014 & Insured’s id, area-code, gender, etc. & \acrshort{rnn} & Accuracy, total error & Python \\ \hline
\cite{Sohangir_2018} & StockTwits  & 2015-2016 & Sentences, StockTwits messages & Doc2vec, \acrshort{cnn} & Accuracy, precision, recall, f-measure, \acrshort{auc} & Python, Tensorflow \\ \hline

\label{table:financial_text_mining_3}
\end{longtable}
\endgroup

In \cite{Ying_2017}, individual social security payment types (paid, unpaid, repaid, transferred) were classified and predicted using \gls{lstm}, \gls{hmm} and \gls{svm}. Sohangir et al. \cite{Sohangir_2018} used two neural network models (doc2Vec, \gls{cnn}) to find the top authors in StockTwits messages and to classify the authors as expert or non-expert for author classification purposes.

In \cite{Costa_2016},  the character sequences in financial transactions and the responses from the other side was used to detect if the transaction was fraud or not with \gls{lstm}. Wang et al. \cite{Wang_2018} used text mining and \gls{dnn} models to  detect automobile insurance fraud.

In \cite{Ronnqvist_2015}, the news semantics were extracted by the word sequence learning, bank stress was determined and classified with the associated events. Day et al. \cite{Day_2016} used financial sentiment analysis using text mining and \gls{dnn} for stock algorithmic trading. 

Cerchiello et al. \cite{Cerchiello_2017} used the fundamental data and text mining from the financial news (Reuters) to classify the bank distress. In \cite{R_nnqvist_2017}, the bank distress was identified by extracting the data from the financial news through text mining. The proposed method used \gls{dfnn} on semantic sentence vectors to classify if there was an event or not. 

\subsection{Theoretical or Conceptual Studies}
There were a number of research papers that \hl{were either} focused on the theoretical concepts of finance or the conceptual designs without model implementation phases; however they still provided valuable information, so we decided to include them in our survey. In Table~\ref{table:theory_or_conceptual_study}, these studies were tabulated according to their topic of interest.

In \cite{Sokolov_2017}, the connection between deep \glspl{ae} and \gls{svd} were discussed and compared using stocks from \gls{ibb} index and the stock of Amgen Inc. Bouchti et al. \cite{Bouchti_2017} \hl{explained} the details of \gls{drl} and mentioned that \gls{drl} could be used for fraud detection/risk management in banking.

\begingroup
\footnotesize
\fontsize{7}{9}\selectfont
\begin{longtable}{
                p{0.03\linewidth}
                p{0.18\linewidth}
                p{0.10\linewidth}
                p{0.18\linewidth}
                p{0.07\linewidth}
                p{0.07\linewidth}
                p{0.13\linewidth}
                }

\caption{Other - Theoretical or Conceptual Studies}\\
\\

\hline
\textbf{Art.}                                                          &
\textbf{SubTopic}                                                          & 
\textbf{IsTimeSeries?}                                              & 
\textbf{Data Set}                                                       & 
\textbf{Period}                                                    & 
\textbf{Feature Set}                                                    & 
\textbf{Method}                                                     \\ 
\hline
\endhead

\cite{Sokolov_2017} & Analysis of \acrshort{ae}, \acrshort{svd} & Yes & Selected stocks from the \acrshort{ibb} index and stock of Amgen Inc. & 2012-2014 & Price data
 & \acrshort{ae}, \acrshort{svd} \\ \hline
\cite{Bouchti_2017} & Fraud Detection in Banking & No & Risk Management / Fraud Detection & - & - & \acrshort{drl} \\ \hline

\label{table:theory_or_conceptual_study}
\end{longtable}
\endgroup

\subsection{Other Financial Applications}
Finally, there were some research papers which did not fit into any of the previously covered topics. Their data set and intended output \hl{were} different than most \hl{of the} other studies focused in this survey. These studies include social security payment classification, bank telemarketing success prediction, hardware solutions for faster financial transaction processing, etc. There were some anomaly detection implementations like tax evasion, money laundering that could have \hl{been} included in this group; however we decided to \hl{cover them in a different subsection, fraud detection.}  Table~\ref{table:other_financial_applications} shows all these aforementioned studies with their differences.  

Dixon et al. \cite{Dixon_2015} used Intel Xeon Phi to speedup the price movement direction prediction problem using \gls{dffn}. The main contribution of the study was the increase in the speed of processing. Alberg et al. \cite{Alberg_2017} used several company financials data (fundamental data) and price together to predict the next period's company financials data. Kim et al. \cite{Kim_2015} used \gls{cnn} for predicting the success of bank telemarketing. In their study, they used the phone calls of the bank marketing data and 16 finance-related attributes. Lee et al. \cite{Lee_2017} used technical indicators and patent information to estimate the revenue and profit for the corporates using \gls{rbm} based \gls{dbn}, \gls{ffnn} and \gls{svr}. 

Ying et al.\cite{Ying_2017} classified and predicted individual social security payment types (paid, unpaid, repaid, transferred) using \gls{lstm}, \gls{hmm} and \gls{svm}. Li et al. \cite{Li_2014} proposed a deep learning-based sentiment analysis method to identify the top sellers in the underground economy. Jeong et al. \cite{Jeong_2019} combined deep Q-learning and deep \gls{nn} to implement \hl{a model to} solve three separate problems: Increasing profit in a market, prediction of the number of shares to trade, and preventing overfitting with insufficient financial data.

\begingroup
\footnotesize

\fontsize{7}{9}\selectfont
\begin{longtable}{
                p{0.03\linewidth}
                p{0.12\linewidth}
                p{0.18\linewidth}
                p{0.08\linewidth}
                p{0.15\linewidth}
                p{0.10\linewidth}
                p{0.10\linewidth}
                p{0.07\linewidth}
                }

\caption{Other Financial Applications}\\
\\

\hline
\textbf{Art.}                                                          & 
\textbf{Subtopic}                                                       &
\textbf{Data Set}                                                       & 
\textbf{Period}                                                    & 
\textbf{Feature Set}                                                    & 
\textbf{Method}                                                 & 
\textbf{Performance Criteria}                                           & 
\textbf{Env.}                                                      \\ 
\hline
\endhead

\cite{Jeong_2019} & Improving trading decisions & \acrshort{sp500}, \acrshort{kospi}, \acrshort{hsi}, and EuroStoxx50 & 1987-2017 & 200-days stock price & Deep Q-Learning and \acrshort{dnn} & Total profit,  Correlation & - \\ \hline
\cite{Li_2014} & Identifying Top Sellers In Underground Economy & Forums data & 2004-2013 & Sentences and keywords & Recursive neural tensor networks & Precision, recall, f-measure & - \\ \hline
\cite{Ying_2017} & Predicting Social Ins. Payment Behavior & Taiwan’s National Pension Insurance & 2008-2014 & Insured’s id, area-code, gender, etc. & \acrshort{rnn} & Accuracy, total error & Python \\ \hline

\cite{Dixon_2015} & Speedup
 & 45 \acrshort{cme}  listed commodity and FX futures & 1991-2014 & Price data & \acrshort{dnn} & - & - \\ \hline
\cite{Alberg_2017} & Forecasting Fundamentals & Stocks in \acrshort{nyse}, \acrshort{nasdaq} or \acrshort{amex} exchanges & 1970-2017 & 16 fundamental features from balance sheet & \acrshort{mlp}, \acrshort{lfm} & \acrshort{mse}, Compount annual return, \acrshort{sr} & - \\ \hline
\cite{Kim_2015} & Predicting Bank Telemarketing & Phone calls of bank marketing data  & 2008-2010 & 16 finance-related attributes & \acrshort{cnn} & Accuracy & - \\ \hline
\cite{Lee_2017} & Corporate Performance Prediction & 22 pharmaceutical companies data in US stock market  & 2000-2015 & 11 financial and 4 patent indicator & \acrshort{rbm}, \acrshort{dbn} & \acrshort{rmse}, profit & - \\ \hline

\label{table:other_financial_applications}
\end{longtable}
\endgroup

\section{Current Snaphot of DL research for Financial Applications}
\label{sec:snapshot}

For the survey, we reviewed 144 papers from various financial application areas. Each paper is analyzed according to its topic, publication type, problem type, method, dataset, feature set and performance criteria. Due to space limitations, we will only provide the general summary statistics indicating the current state of the \gls{dl} for finance research. 

\begin{figure*}[htb]
\centering
\includegraphics[width=4.5in]{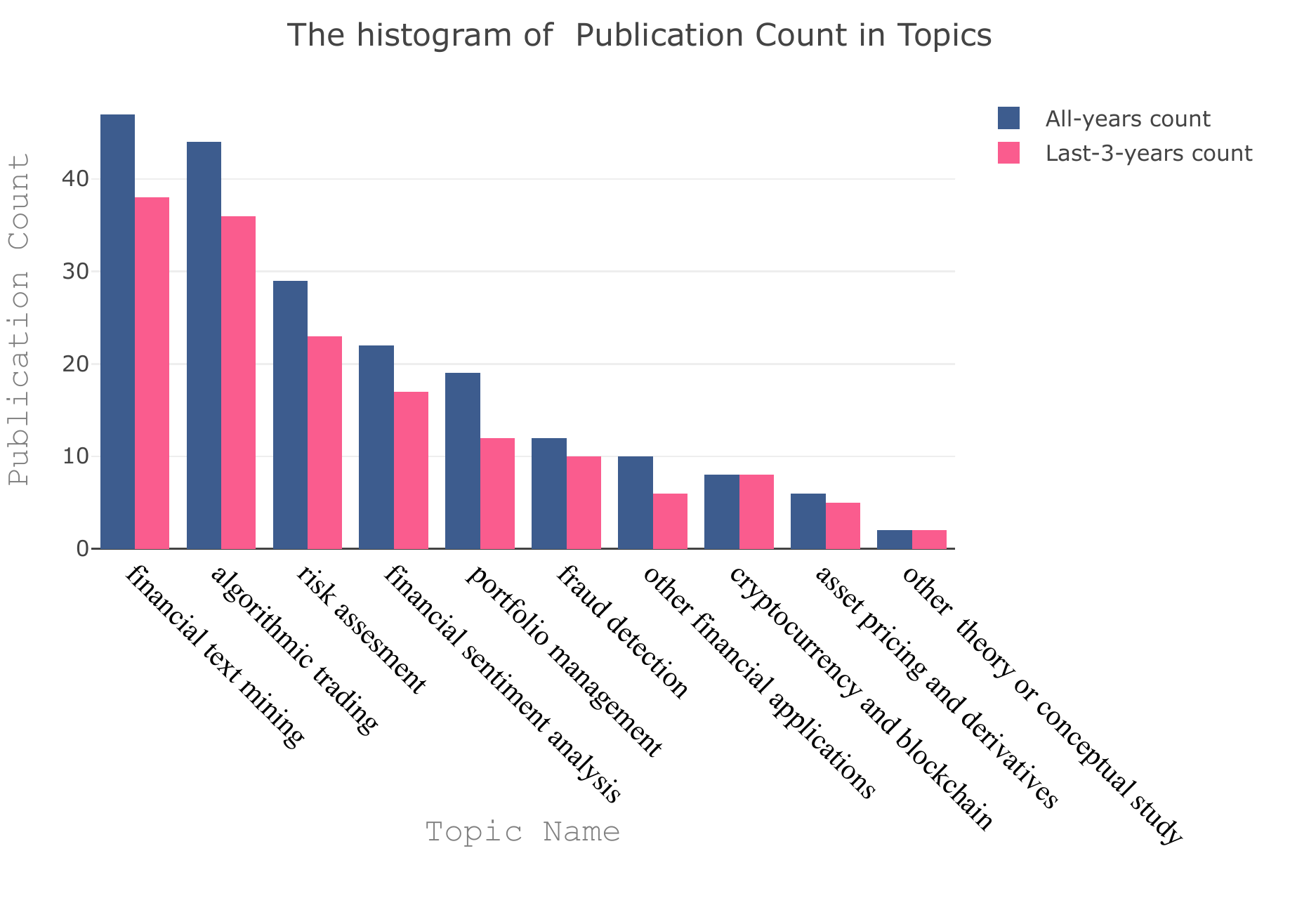}
\caption{The histogram of  Publication Count in Topics}
\label{fig:histogram_of_topic}
\end{figure*}

\begin{figure*}[!htb]
\centering
\includegraphics[width=4.5in]{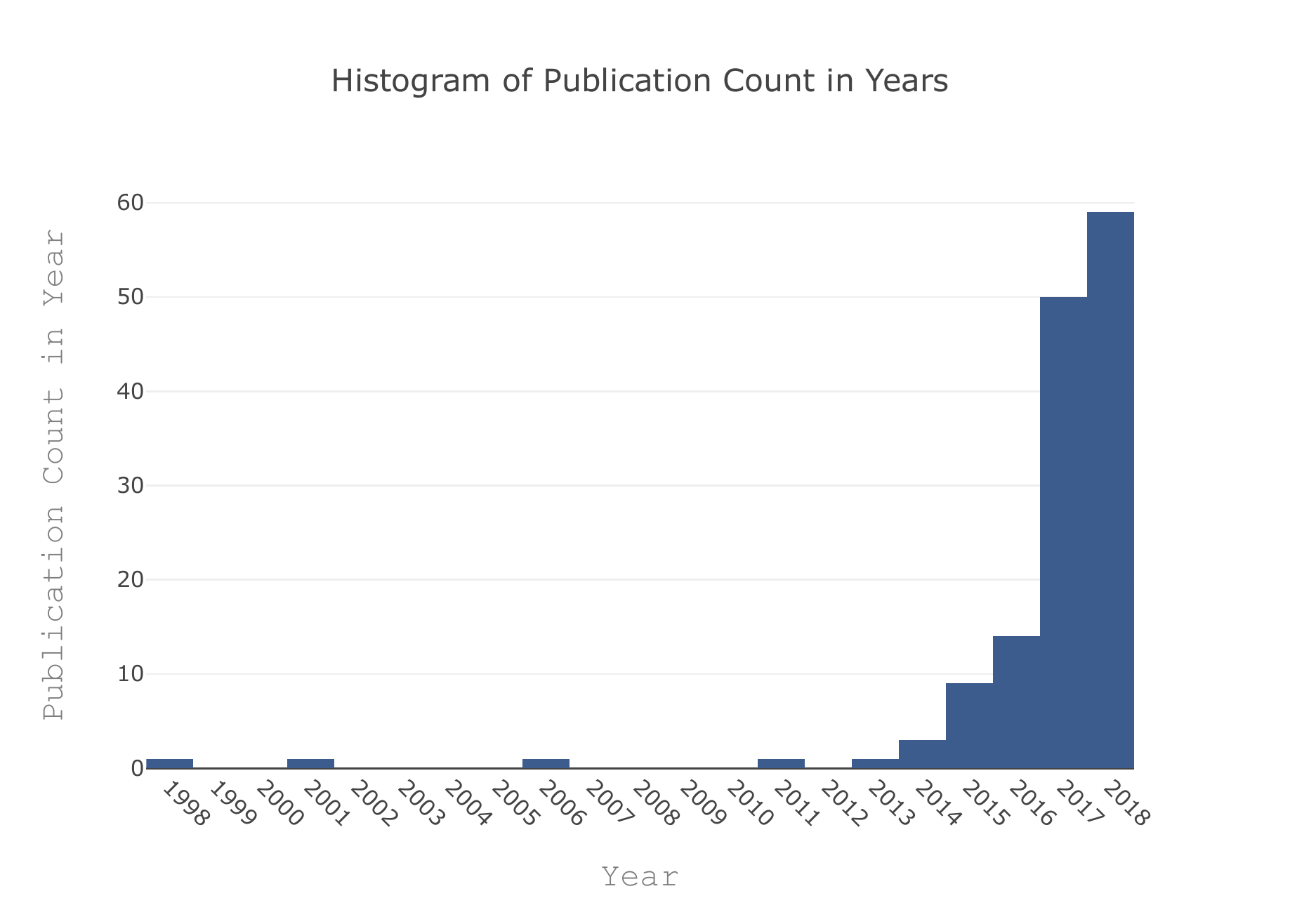}
\caption{The histogram of  Publication Count in Years}
\label{fig:histogram_of_year}
\end{figure*}

\begin{figure*}[!htb]
\centering
\includegraphics[width=4.5in]{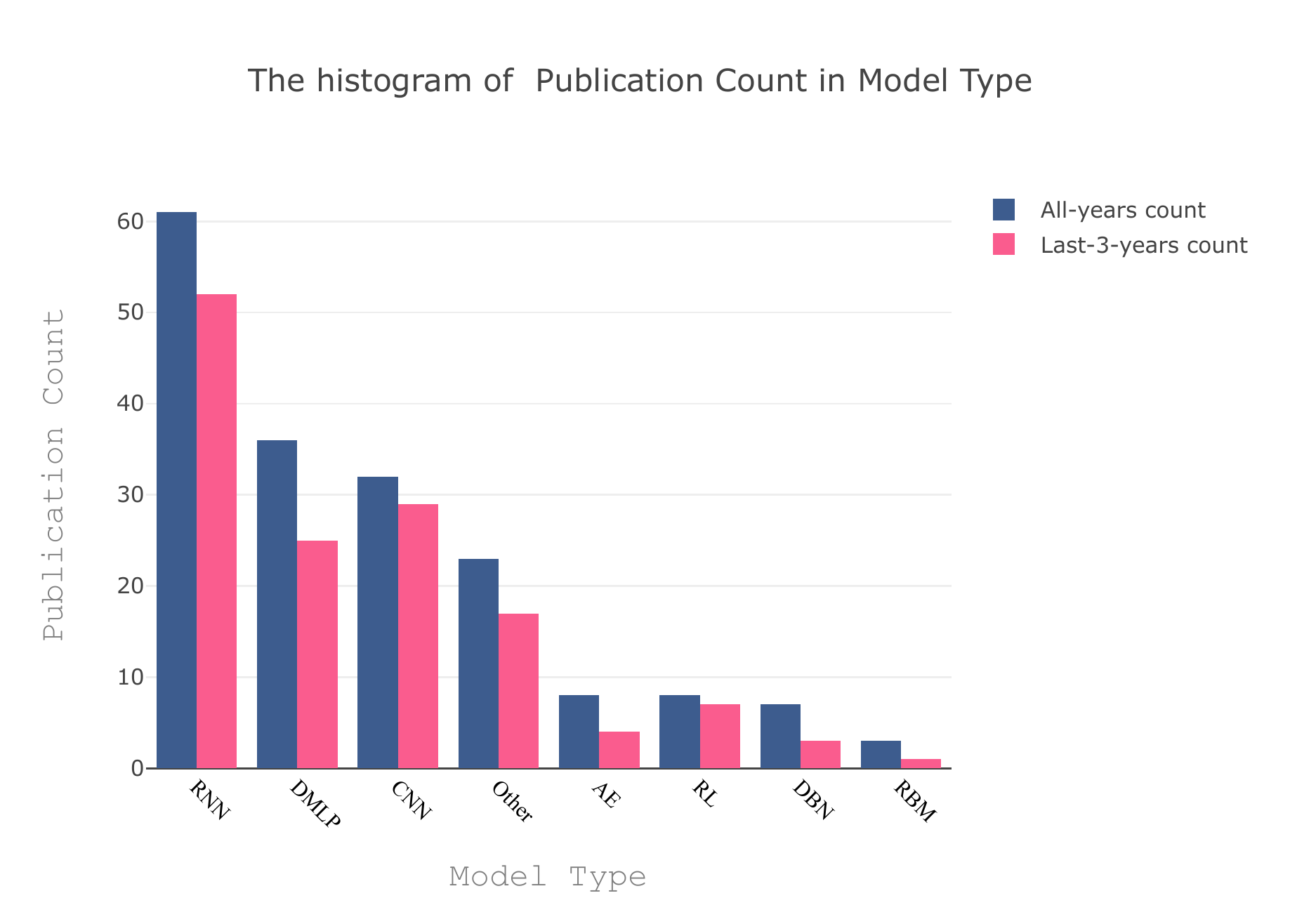}
\caption{The histogram of  Publication Count in Model Types}
\label{fig:histogram_of_model_type}
\end{figure*}

First and foremost, we clustered the various topics within the financial applications research and presented them in Figure \ref{fig:histogram_of_topic}. A quick glance at the figure shows us financial text mining and algorithmic trading are the top two fields that the researchers most worked on followed by risk assessment, sentiment analysis, portfolio management and fraud detection, respectively. The results indicate most of the papers were published within the last 3 years implying the domain is very hot and actively studied. We can also observe these phenomena by analyzing Figure \ref{fig:histogram_of_year}. Also, it is worth to mention that the few papers that were published before 2013 all used \gls{rnn} based models.

When the papers were clustered by the \gls{dl} model type as presented in Figure \ref{fig:histogram_of_model_type}, we observe the dominance of \gls{rnn}, \gls{dmlp} and \gls{cnn} over the remaining models, which might be expected, since these models are the most commonly preferred ones in general \gls{dl} implementations. Meanwhile, \gls{rnn} is a general umbrella model which has several versions including \gls{lstm}, \gls{gru}, etc. Within the \gls{rnn} choice, most of the models actually belonged to \gls{lstm}, which is very popular in time series forecasting or regression problems. It is also used quite often in algorithmic trading. More than 70\% of the \gls{rnn} papers consisted of \gls{lstm} models. 

\begin{figure*}[!htb]
\centering
\includegraphics[width=2.5in]{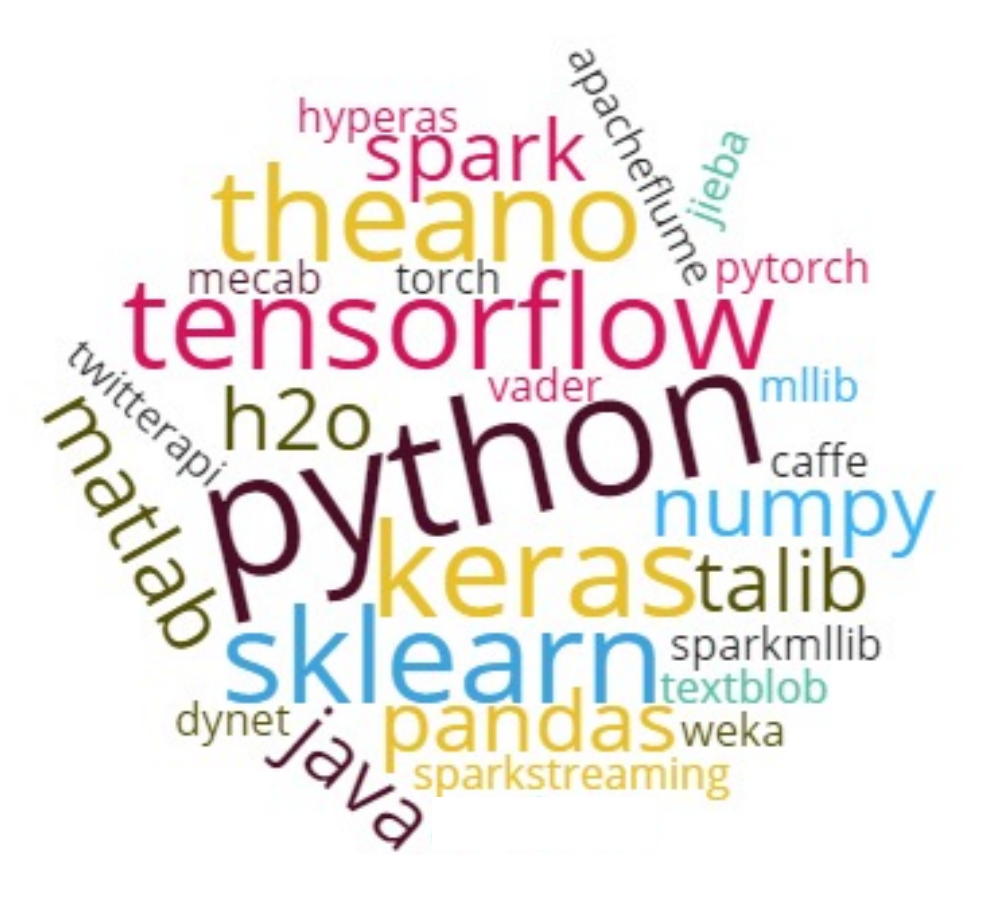}
\caption{Wordcloud of most-used Software, Frameworks, Environments}
\label{fig:wordcloud}
\end{figure*}

\begin{figure*}[!htb]
    \centering
    \begin{subfigure}{.5\textwidth}
        \centering
        \includegraphics[width=.99\linewidth]{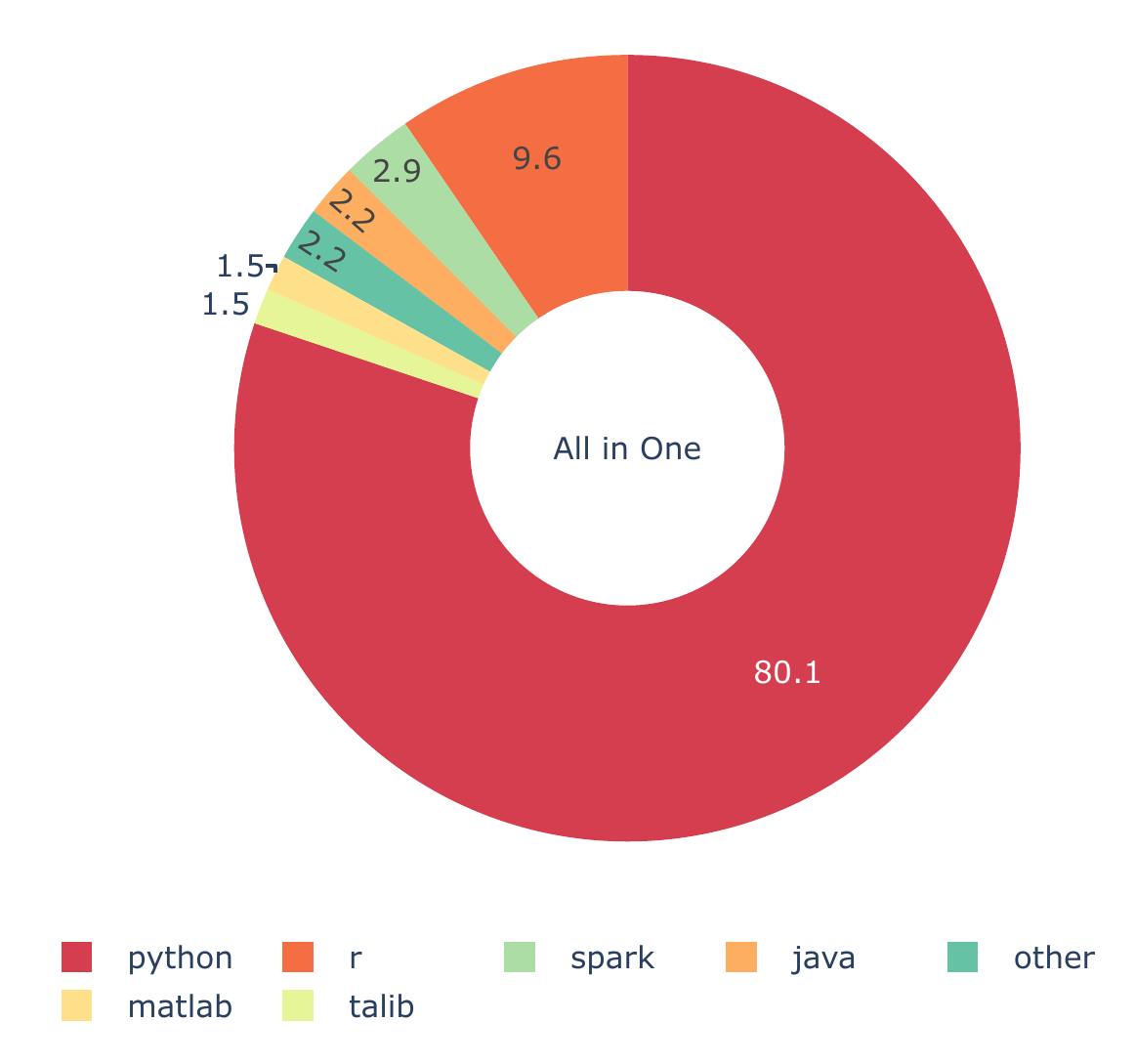}
        \caption{Preferred Development Environments}
        \label{fig:platform_piechart_all}
    \end{subfigure}%
    \begin{subfigure}{.5\textwidth}
        \centering
        \includegraphics[width=.99\linewidth]{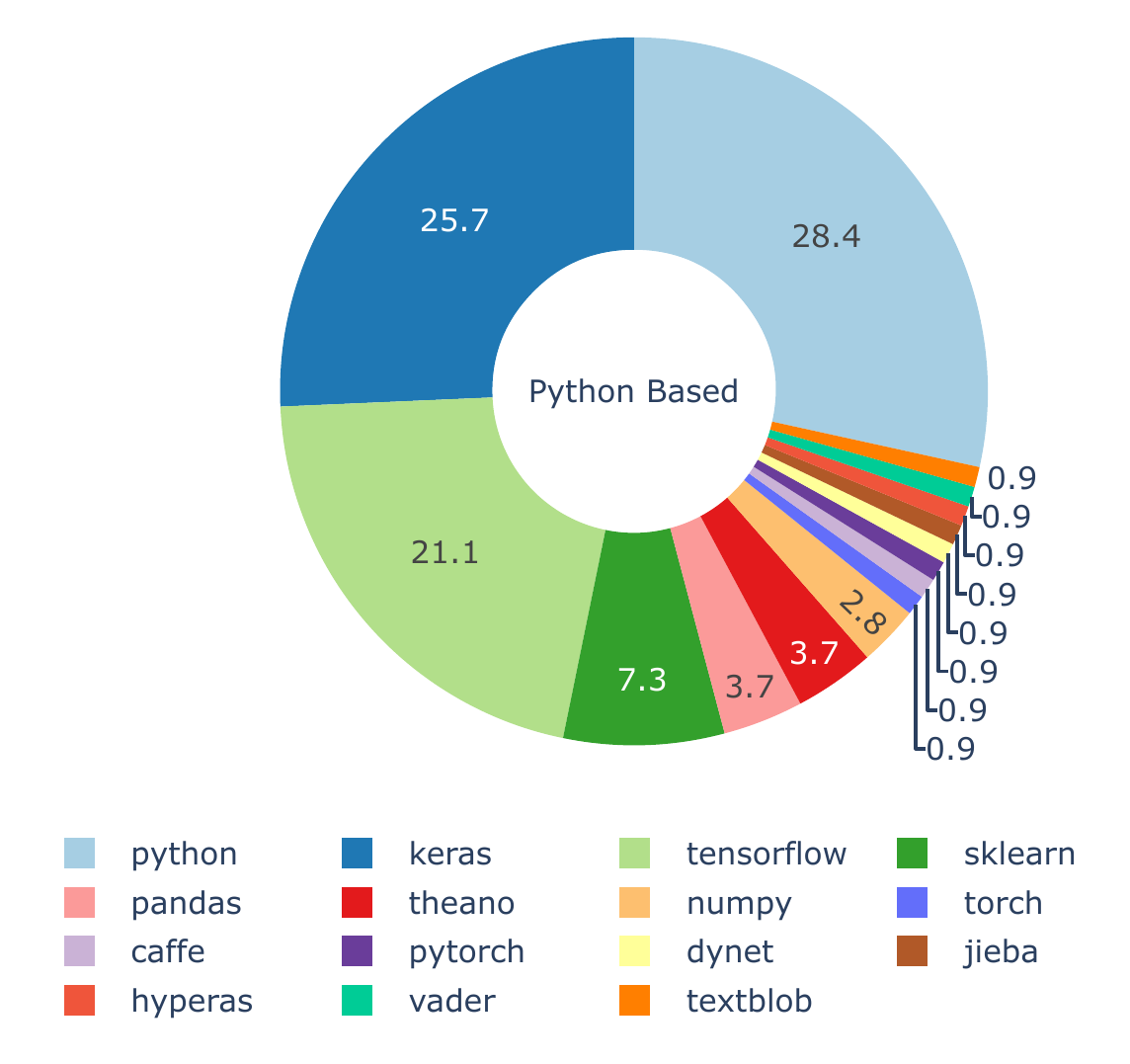}
        \caption{Preferred Python Libraries}
        \label{fig:platform_piechart_py}
    \end{subfigure}
    \caption{Distribution of Preferred Environments}
    \label{fig:platform_piechart}
\end{figure*}

Figure \ref{fig:wordcloud} \hl{presents the commonly used software and frameworks for} \gls{dl} \hl{model implementations through Wordcloud whereas Figure } \ref{fig:platform_piechart} \hl{provides the details about the development environments. The left chart (Figure } \ref{fig:platform_piechart_all}\hl{) presents the high level view where Python had the lion's share with 80\% over R (with 10\%) and the other languages. The chart on the right (Figure } \ref{fig:platform_piechart_py}\hl{) provides the details about how the developers are using Python through different libraries and frameworks.} 

Meanwhile, \gls{dmlp} generally fits well for classification problems; hence it is a common choice for most of the financial application areas. However, since it is a natural extension of its shallow counterpart \gls{mlp}, it has a longer history than the other \gls{dl} models.

\begin{figure*}[!htb]
\centering
\includegraphics[width=\linewidth]{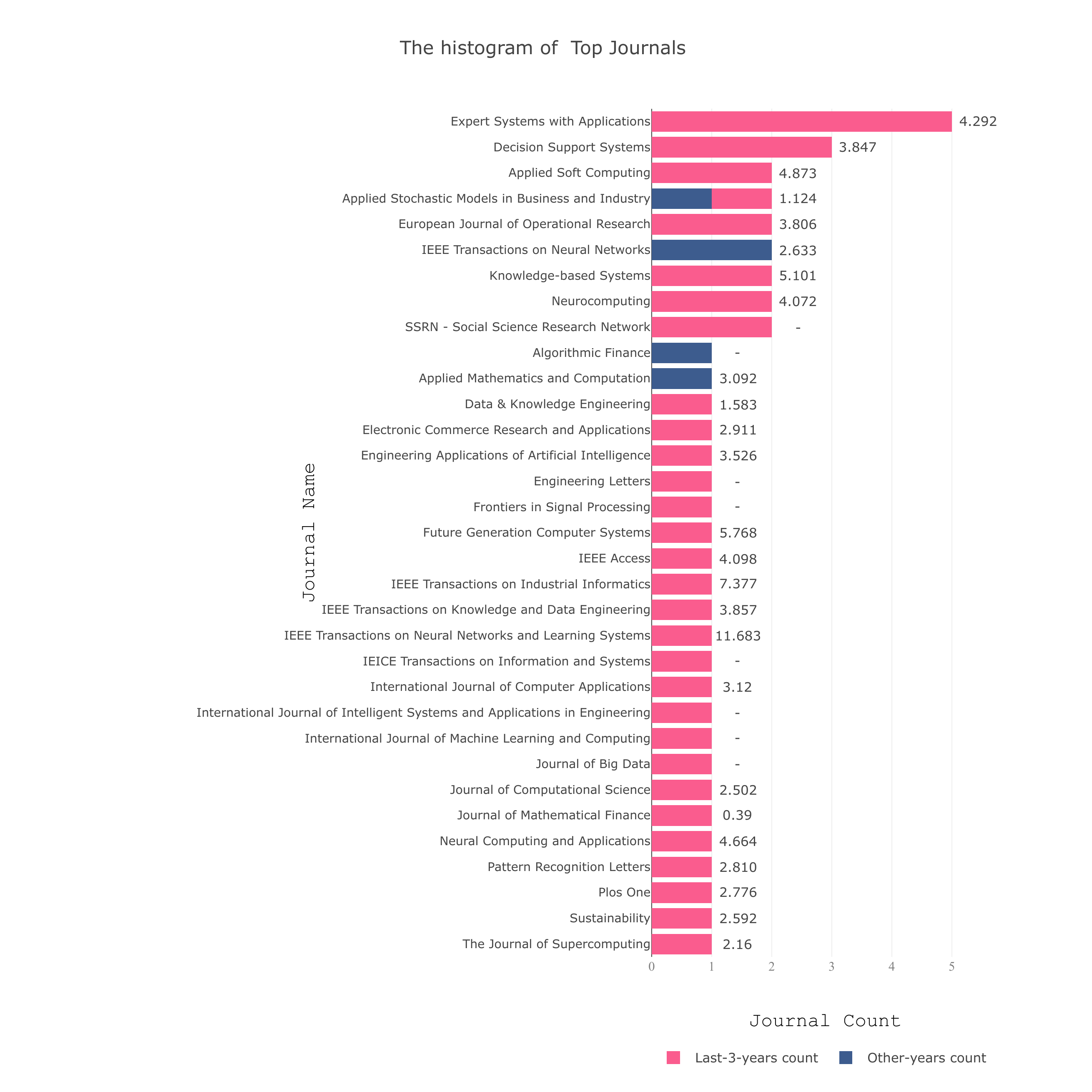}
\caption{Top Journals - corresponding numbers next to the bar graph are representing the impact factor of the journals}
\label{fig:top_journals}
\end{figure*}

\gls{cnn} started getting more attention lately since most of the implementations appeared within the last 3 years. Careful analysis of \gls{cnn} papers indicates that a recent trend of representing financial data with a 2-D image view in order to utilize \gls{cnn} is growing. Hence \gls{cnn} based models might overpass the other models in the future. It actually passed \gls{dmlp} for the last 3 years.

The top journals are tabulated in Fig~\ref{fig:top_journals}. The journals with the most published papers in the last 3 years include Expert Systems with Applications, Decision Support Systems, Applied Soft Computing, Neurocomputing, Knowledge-based Systems and European Journal of Operational Research.


\section{Discussion and Open Issues}
\label{sec:discussion}

After reviewing all the publications based on the selected criteria explained in the previous section, we wanted to provide our findings of the current state-of-the-art situation. Our discussions are categorized by the \gls{dl} models and implementation topics.

\subsection{Discussions on DL Models}

It is possible to claim that \gls{lstm} is the dominant \gls{dl} model that is preferred by most researchers, due to its well-established structure for financial time series data forecasting. Most of the financial implementations have time-varying data representations requiring regression-type approaches which fits very well for \gls{lstm} and its derivatives due to their easy adaptations to the problems. As long as the temporal nature of the financial data remains, \gls{lstm} and its related family models will maintain their popularities.

Meanwhile, \gls{cnn} based models started getting more traction among researchers in the last two years. Unlike \gls{lstm}, \gls{cnn} works better for classification problems and is more suitable for either non-time varying or static data representations. However, since most financial data is time-varying, under normal circumstances, \gls{cnn} is not the natural choice for financial applications. However, in some independent studies, the researchers performed an innovative transformation of 1-D time-varying financial data into 2-D mostly stationary image-like data to be able to utilize the power of \gls{cnn} through adaptive filtering and implicit dimensionality reduction. This novel approach seems working remarkably well in complex financial patterns regardless of the application area. In the future, more examples of such implementations might be more common; only time will tell.

Another model that has a rising interest is \gls{drl} based implementations; in particular, the ones coupled with agent-based modelling. Even though algorithmic trading is the most preferred implementation area for such models, it is possible to develop the working structures for any problem type.

Careful analyses of the reviews indicate in most of the papers hybrid models are preferred over native models for better accomplishments. A lot of researchers configure the topologies and network parameters for achieving higher performance. However, there is also the danger of creating more complex hybrid models that are not easy to build, and their interpretation also might be difficult.

Through the performance evaluation results, it is possible to claim that in general terms, \gls{dl} models outperform \gls{ml} counterparts when working on the same problems. \gls{dl} problems also have the advantage of being able to work on larger amount of data. With the growing expansion of open-source \gls{dl} libraries and frameworks (Figure \ref{fig:wordcloud}), \gls{dl} model building and development process is easier than ever. And this phenomena is also supported by the increasing interest in adapting \gls{dl} models into all areas of finance which can be observed from Figure  \ref{fig:histogram_of_year}. 

 Also it is worth to mention that, besides the outperformance of \gls{dl} models over \gls{ml}, the performance evaluation results are improving every year relatively, even though it is very difficult to explicitly quantifty the amount of improvement. The improvements are most notable in trend prediction based algo-trading implementations and text-mining studies due to deeper and/or more versatile networks and new innovative model developments. This is also reflected through the increasing number of published papers year over year.

\subsection{Discussions on Implementation Areas}

Price/trend prediction and Algo-trading models have the most interest among all financial applications that use \gls{dl} models in their implementations. Risk assessment and portfolio management have always been popular within the \gls{ml} community, and it looks like this is also valid for \gls{dl} researchers.

Even though broad interest in \gls{dl} models is on the rise, financial text mining is particularly getting more attention than most of the other financial applications. The streaming flow of financial news, tweets, statements, blogs opened up a whole new world for the financial community allowing them to build better and more versatile prediction and evaluation models integrating numerical and textual data. Meanwhile, the general approach nowadays is to combine text mining with financial sentiment analysis. With that, it is reasonable to assume higher performance will be achieved. A lot of researchers started working on that particular application area. It is quite probable that the next generation of outperforming implementations will be based on models that can successfully integrate text mining with quantified numerical data.

These days, one other hot area \hl{within the} \gls{dl} research is the cryptocurrencies. We can also include blockchain research to that, even though it is not necessarily directly related to cryptocurrencies, but generally used together in most implementations. Cryptocurrency price prediction has the most attraction within the field, but since the topic is fairly new, more studies and implementations will probably keep pouring in due to the high expectations and promising rewards.

\subsection{Open Issues and Future Work}

When we try to extrapolate the current state of research and the achieved accomplishments into the future, a few areas of interests stand out. We will try to elaborate on them and provide a pathway for what can be done or needs to be done within the following few years. We will try to sort out our opinions by analyzing them through the model development and research topic point of view.
 
\subsubsection{Model Development Perspective}

We have already mentioned the growing attention on the adaptation of 2-D \gls{cnn} implementations for various financial application areas. This particular technique looks promising and provides opportunities. It would be beneficial to further explore the possibilities using that approach in different problems. The playfield is still wide open.

Graph \gls{cnn} is another model that is closely related but still showing some discrepancies. It has not been used much, only one study was published that relates graph-\gls{cnn} with financial applications. However, versatile transformations of financial data into graphs, integrating sentiment analysis through graph representations and constructing different models can create opportunities for researchers to build better performing financial applications. 

There are also recently developed \gls{dl} models, like \gls{gan}, Capsule networks, etc. that can also provide viable alternatives to existing implementations. They have started showing up in various non-financial studies, however to the best of our knowledge, no known implementation of such kind for financial applications exists. It might open up a new window of opportunities for financial researchers and practitioners. In addition to such new models, innovative paradigms like transfer learning, one-shot learning can be tested within the environment.

Since financial text mining is overtaking the other topics in an accelerated fashion, new data models like Stock2Vec \cite{Dang_2018} can be enhanced for better and more representative models. In addition, \gls{nlp} based ensemble models or more integration of data semantics into the picture can increase the accuracy of the existing models.

Finally, according to our observations, hybrid models are preferred more over the native or standalone models in most studies. This trend will likely continue, however, researchers need to introduce more versatile, sometimes unconventional models for better results. Hybrid models integrating various simple \gls{dl} layers like cascaded \gls{cnn}-\gls{lstm} blocks can have better outcomes since ensembling spatial and temporal information together in a novel way might be an important milestone for researchers seeking for "alpha" in their models. 

\subsubsection{Implementation Perspective}

As far as the application areas are concerned, the usual suspects, algorithmic trading, portfolio management and risk assessment will probably continue on their dominance within the financial research arena in the foreseeable future. Meanwhile, some new shining stars started getting more attention, not only because they represent fairly new research opportunities, but also their forecasted impact on the financial world is noteworthy.

Cryptocurrencies and blockchain technology are among these new research areas. Hence, it is worthwhile to explore the possibilities that these new fields will bring. It will be a while before any of these technologies become widely accepted industry standard, however, that is the sole reason why it provides a great opportunity for the researchers to shape the future of the financial world with new innovative models and hoping that the rest of the world will follow their footsteps.

Another area that can benefit from more innovative models is portfolio management. Robo-advisory systems are on the rise throughout the world and these systems depend on high performing automated decision support systems. Since \gls{dl} models fit well to that description, it would be logical to assume the utilization of \gls{dl} implementations will increase in the coming years. As such, the corresponding quant funds will be very interested in the achievements that the \gls{dl} researchers can offer for the financial community.  This might require integrating learning and optimization models together for better-performing systems. Hence, ensemble models that can successfully mix \gls{ec} and \gls{dl} components might be what the industry is anticipating for the immediate future. This might also result in new research opportunities.

Yet, one other research area that is generally avoided by soft computing and \gls{dl} researchers is the financial derivatives market. Even though there are many different products that exist on the market, the corresponding \gls{dl} research is very scarce. However, for professionals working in the finance industry, these products actually provide incredible flexibilities ranging from hedging their investments to implementing leveraged transactions with minimized risk. Even though, opportunities exist for \gls{dl} researchers, there was not a broad interest in the topic, since there are only a handful of studies for the derivatives market. Option strategy optimization, futures trading, option pricing, arbitrage trading can be among the areas that might benefit from \gls{dl} research. 

Sentiment analysis, text mining, risk adjusted asset pricing are some of the other implementation areas that attract researchers but not yet fully utilized. It is quite probable we will see more papers in these fields in the near future.

Last, but not least, \gls{hft} is one area that has not benefitted from the advancements in \gls{ml} research to its full potential yet. Since \gls{hft} requires lightning-fast transaction processing, the statistical learning model that is embedded into such trading systems must not introduce any extra latency to the existing system. This necessitates careful planning and modelling of such models. For that purpose, \gls{dl} models embedded within the \gls{gpu} or \gls{fpga} based hardware solutions can be studied. The hardware aspects of \gls{dl} implementations are generally omitted in almost all studies, but as stated above, there might be opportunities also in that field.

\subsubsection{Suggestions for Future Research}

Careful analyses of Figures  \ref{fig:histogram_of_topic} and \ref{fig:histogram_of_year} indicate the rising overall appetite for applied \gls{dl} research for finance. Even though the interest is broad, some areas like cryptocurrency and block chain studies might get more attention compared to other areas.

With respect to the promising outlook in text mining and financial sentiment analysis, we believe behavioral finance is also a fairly untouched research area that hides a lot of opportunities within. There is a lack of research work published on behavioral finance using \gls{dl} models. This might be mainly due to the difficulties of quantifying the inputs and outputs of behavioral finance research to be used with \gls{dl} models. However, new advancements in text mining, \gls{nlp}, semantics combined with agent-based computational finance can open up huge opportunities in that field. We would encourage researchers to look further into this for a possible implementation area as it currently seems to be wide open for new studies.

\subsection{Responses to our Initial Research Questions}

At this point, since we gathered and processed all the information we need, we are ready to provide answers to our initially stated research questions. The questions and our corresponding answers according to our survey are as follows: 

\begin{itemize}
\item {What financial application areas are of interest to \gls{dl} community?}

Response: Financial text mining, Algo-trading, risk assessments, sentiment analysis, portfolio management and fraud detection are among the most studied areas of finance research. (Please check Figure \ref{fig:histogram_of_topic}) 

\item {How mature is the existing research in each of these application areas?}

Response: Even though \gls{dl} models already had better achievements compared to traditional counterparts in almost all areas, the overall interest is still on the rise in all research areas.

\item {What are the areas that have promising potentials from an academic/industrial
research perspective?}

Response: Cryptocurrencies, blockchain, behavioral finance, \gls{hft} and derivatives market have promising potentials for research.

\item {Which \gls{dl} models are preferred (and more successful) in different applications?}

Response: \gls{rnn} based models (in particular \gls{lstm}), \gls{cnn} and \gls{dmlp} have been used extensively in implementations. From what we \hl{have}  encountered, \gls{lstm} is more successful and preferred in time-series forecasting, whereas \gls{dmlp} and \gls{cnn} are better suited to applications requiring classification. 

\item {How do \gls{dl} models pare against traditional soft computing / \gls{ml} techniques?}

Response: In most of the studies, \gls{dl} models performed better than their \gls{ml} counterparts. There were a few occasions where \gls{ml} had comparable or even better solutions, however the general tendency is the outperformance of the \gls{dl} methods.

\item {What is the future direction for \gls{dl} research in Finance?}

Response: Hybrid models based on Spatio-temporal data representations, \gls{nlp}, semantics and text mining-based models might become more important in the near future.

\end{itemize}

\section{Conclusions}
\label{sec:conclusions}

The financial industry and academia have started realizing the potentials of \gls{dl} in various application areas. The number of research work keeps on increasing every year with an accelerated fashion. However, we are just in the early years of this new era, more studies will be implemented and new models will keep pouring in. In this survey, we wanted to highlight the state-of-the-art \gls{dl} research for the financial applications. We not only provided a snapshot of the existing research status but also tried to identify the future roadway for intended researchers. Our findings indicate there are incredible opportunities within the field and it looks like they will not disappear anytime soon. So, we encourage the researchers that are interested in the area to start exploring.

\section{Acknowledgement}

This work is supported by the Scientific and Technological Research Council of Turkey (TUBITAK) grant no 215E248.



\renewcommand{\glsgroupskip}{}
\printglossaries

\clearpage

\bibliographystyle{unsrtnat_wo_doi}
\bibliography{bibdatabase}

\end{document}